\documentclass[usenatbib,usegraphicx,useAMS]{mn2e}
\usepackage[T1]{fontenc}
\usepackage{natbib}
\usepackage{amssymb}
\usepackage{array}
\usepackage{color}
\usepackage[normalem]{ulem}
\usepackage{aas_macros}
\begin{document}
\title[A sensitivity analysis of the WTS]{A sensitivity analysis of
  the WFCAM Transit Survey for short-period giant planets around M
  dwarfs}  
\author[G.\ Kov\'acs et al.]{G\'abor Kov\'acs$^1$, S.\ Hodgkin$^1$,
  B.\ Sip\H{o}cz$^2$, D.\ Pinfield$^2$, \newauthor D.\ Barrado$^3$,
  J.\ Birkby$^4$, M.\ Cappetta$^5$, P.\ Cruz$^3$, 
  J.\ Koppenhoefer$^5$, E.\ Mart\'\i n$^6$, \newauthor
  F.\ Murgas$^6$, B.\ Nefs$^4$, R.\ Saglia$^5$, J.\ Zendejas$^5$\\
  $^1$Institute of Astronomy, University of Cambridge, Madingley Road,
  Cambridge, CB3 0HA, UK\\
  $^2$Centre for Astrophysics Research, University of Hertfordshire,
  College Lane, Hatfield, AL10 9AB, UK\\
  $^3$Centro de Astrobiolog\'\i a, Instituto Nacional de T\'ecnica Aeroespacial, 28850 Torrej\'on de Ardoz, Madrid, Spain\\
  $^4$Leiden Observatory, Universiteit Leiden, Niels Bohrweg 2, NL-2333 CA Leiden, The Netherlands\\
  $^5$Max-Planck-Institut f\"ur extraterrestrische Physik, Giessenbachstra\ss e, 85748 Garching, Germany\\
  $^6$Instituto de Astrof\'\i sica de Canarias,C/ V\'\i a L\'actea s/n, E38205 La Laguna (Tenerife), Spain\\
} \bibliographystyle{mn2e}
\maketitle
\defcitealias{Howard_2012}{H12}
\begin{abstract}
  The WFCAM Transit Survey (WTS) is a near-infrared transit survey
  running on the United Kingdom Infrared Telescope (UKIRT), designed
  to discover planets around M dwarfs. The WTS acts as a poor-seeing
  backup programme for the telescope, and represents the first
  dedicated wide-field near-infrared transit survey. Observations
  began in 2007 gathering J-band photometric observations in four
  (seasonal) fields. In this paper we present an analysis of the first
  of the WTS fields, covering an area of 1.6 square degrees. We
  describe the observing strategy of the WTS and the processing of the
  data to generate lightcurves. We describe the basic
    properties of our photometric data, and measure our sensitivity
    based on 950 observations. We show that the photometry reaches a
  precision of $\sim 4$ mmag for the brightest unsaturated stars in
  lightcurves spanning almost 3 years. Optical (SDSS $griz$) and
  near-infrared (UKIRT $ZYJHK$) photometry is used to classify the
  target sample of 4600 M dwarfs with J magnitudes in the range
  11--17. Most have spectral-types in the range M0--M2. We conduct
  Monte Carlo transit injection and detection simulations for short
  period ($<$10 day) Jupiter- and Neptune-sized planets to
  characterize the sensitivity of the survey. We investigate the
  recovery rate as a function of period and magnitude for 4
  hypothetical star-planet cases: M0--2$+$Jupiter, M2--4$+$Jupiter,
  M0--2$+$Neptune, M2--4$+$Neptune. We find that the WTS lightcurves
  are very sensitive to the presence of Jupiter-sized short-period
  transiting planets around M dwarfs. Hot Neptunes produce a much
  weaker signal and suffer a correspondingly smaller recovery
  fraction. Neptunes can only be reliably recovered with the correct
  period around the rather small sample ($\sim 100$) of the latest M
  dwarfs (M4--M9) in the WTS. The non-detection of a hot-Jupiter
  around an M dwarf by the WFCAM Transit Survey allows us to place an
  upper limit of 1.7--2.0 per cent (at 95 per cent confidence) on
  the planet occurrence rate.
\end{abstract}

\section{Introduction}
M dwarfs are the most numerous stars in our Galaxy
\citep{Chabrier_2003}, and until recently have remained relatively
unexplored as exoplanet hosts. While several hundreds of transiting
exoplanets\footnote{http://exoplanet.eu} have been discovered around
F,G and K dwarfs, only $\sim 70$ such planets are
known around the lower mass M stars. M dwarfs make interesting targets
for many reasons, for example they provide better sensitivity to
smaller (rocky) planets in the habitable zone, and they also provide
strong constraints on planetary formation theories. 
\par The most successful techniques so far used to discover exoplanets
are the radial velocity (RV) and transit methods. A
  planet around an early M2 dwarf can have $\sim$4 times deeper
  transits and $\sim$1.5 times higher RV amplitudes at the same
  orbital period than orbiting a Sun-like star. Both methods also show
  increasing sensitivity towards smaller semi-major axis. Around
solar-type stars, planets in short period orbits (days) are extremely
hot (above 1000K). If M dwarfs harbour planets in equally close-in
orbits, they will probably be more interesting due to their moderate
equilibrium temperatures. In case of later (M4-9) type
  M dwarfs, these planets may fall in the star's habitable zone, and
for rocky planets water can exist as a liquid, while Neptune-like
planets may have water vapour in their atmospheres.
\par M dwarfs are also important for testing planet formation
theories. In the core accretion paradigm, dust particles in the
protoplanetary disk coagulate and form solid
cores. The cores then begin to accumulate gas from the
  disk. When cores reach several Earth masses, gas accumulation can
  speed up significantly. It was first shown by \cite{Laughlin_2004}
that gas giants cannot form easily around low mass stars this way. In
the core accretion process gas giant formation is inhibited by time
scale differences. The protoplanetary disk around a low mass host
dissipates before planetary cores are able to accrete their gas
envelope by entering the runaway gas accretion phase. These planets
remain `failed cores'. \cite{Ida_Lin_2005} found very similar
results. Newer models refine this picture by introducing more detailed
relations between stellar and disk properties. They consider migration
driven by disk-planet interactions (type I-II) \citep{Ida_Lin_2008},
evolution of snow-line location \citep{Kennedy_Kenyon_2008} and
planet-planet scattering \citep{Thommes_2008}. Their conclusions allow
gas giant formation around low mass stars but the predicted frequency
of giants systematically decreases towards lower primary
masses. \cite{Kennedy_Kenyon_2008} give relative ratios for gas giant
planets\footnote{Planet formation theories usually do not provide
  absolute numbers because of free interaction coefficients in their
  formulae.}: a fraction of 1 per cent of low mass stars is predicted
to have at least one giant planet, assuming that this ratio is 6 per
cent for solar mass ones.
\par Due to the failed core outcome, it is not surprising that Neptunes
and rocky planets are predicted to be common around low mass stars in
core accretion models. \cite{Ida_Lin_2005} predict the highest
frequency around M dwarfs for a few Earth mass planets in close-in
orbits ($<$0.05 AU). Later models also firmly support the existence of
these smaller planets \citep{Ida_Lin_2008,Kennedy_Kenyon_2008}.
\par Even if it is hard to form giant planets via core accretion
around the lowest mass stars, gravitational instability models can
produce gas giants on a very short timescale ($\sim
10^3\mathrm{yr}$). They predict that gas giants can form around low
mass primaries as efficiently as around more massive ones assuming
that the protoplanetary disk is sufficiently massive to become
unstable \citep{Boss_2006}.
%
\par \cite[][and references therein]{Wright_2012} give a summary of
occurrence rates of hot giant planets ($T<10$ days) around solar G
type dwarfs. RV studies determined a rate of 0.9--1.5 per cent
\citep{Marcy_2005,Cumming_2008,Mayor_2011,Wright_2012}, while transit
studies have a systematically lower rate (roughly half of this value)
at 0.3--0.5 per cent \citep[][H12 hereafter]{Gould_2006,Howard_2012}.
\par For M dwarfs, recent RV studies support the paucity of giant
planets \citep{Cumming_2008,Johnson_2007,Johnson_2010,Rodler_2012}
though there are confirmed detections both with short- and long-period
orbits. \cite{Johnson_2007} found 3 Jovian planets (with orbital
periods of years) in a sample of 169 K and M dwarfs, in the California
and Carnegie Planet Search data. The planetary occurrence rate for
stars with M$<0.7$M$_\odot$ in their survey is 1.8 per cent, which is
significantly lower than the rate found around more massive hosts (4.2
per cent for Solar-mass stars, 8.9 per cent around higher mass
subgiants). The positive correlation between planet occurrence and
stellar mass remains after metallicity is taken into account, although
at somewhat lower significance. Gravitational microlensing programmes
also detected Jupiter-like giants around M dwarfs
(e.g. \cite{Gould_2010,Batista_2011}) but statistical studies seem to
arrive at different conclusions. Microlensing surveys are more
sensitive to longer period systems than RV (and particularly transit)
surveys. \cite{Gould_2010} analyzed 13 high magnification microlensing
events with 6 planet discoveries around low mass hosts (typically
$0.5M_\oplus$) and derived planet frequencies from a small but
arguably unbiased sample. They compared their planet frequencies to
the \cite{Cumming_2008} RV study. They found that after rescaling with
the snow-line distance to account for their lower stellar masses in
the microlensing case, the planet frequency at high semi-major axes is
consistent with the distribution from the RV study extrapolated to
their long orbits. They found a planet fraction at
semi-major axes beyond the snow line to be 8 times higher than at 0.3
AU. Considering that hot giants in RV discoveries (around solar type
stars) are thought to have migrated large distances into their short
orbits, they conclude that giant planets discovered at high semi-major
axes around low mass hosts do not migrate very far. Their study
suggests that rather than the formation characteristics, the migration
of gas giants may be different in the low mass case.
\par Focusing again on giant planets with short orbital periods around M
dwarfs, then as of writing, there is only one confirmed detection. The
Kepler Mission \citep{Kepler}, includes a sample of 1086 low-mass targets
(in Q2) \citep[][\citetalias{Howard_2012}]{Borucki_2011} and one
confirmed hot Jupiter (P=2.45 days) around an early M dwarf host (KOI-254)
\citep{Johnson_2012}.  Unfortunately, this object was not included in
the statistical analysis of \citetalias{Howard_2012}
study.
\par For transit surveys, sufficiently bright M dwarfs make good
targets because of their smaller stellar radius.  A Neptune-like
object in front of a smaller host can produce a similar photometric
dip ($\sim1$ per cent) as a Jupiter-radius planet orbiting a Sun-like
star. Ground based surveys are therefore potentially sensitive to
planets around M dwarfs that could not be detected around earlier type
stars with typical ground-based precision. On the other hand, M dwarfs
provide a more demanding technical challenge.  They are intrinsically
faint and their spectral energy distribution peaks in the near
infrared. Their faintness at optical wavelengths also makes
spectroscopic follow-up observations difficult. Measuring photometry
in the infrared helps, but introduces a higher sky background level.
It is also worth pointing out that the intrinsic variability of M
dwarfs (flaring, spots) could further reduce the ease of discovering
transiting systems.
\par One approach is to target a large number of M dwarfs, with a
wide-field camera equipped with optical or near-infrared detectors. An
alternative approach is to target individual brighter M dwarfs in the
optical deploying several small telescopes. Up till now, only 2
transiting planets have been discovered around (bright) M stars
\citep{Gillon_2007,Charbonneau_2009} by {\em ground based} transit surveys.
\par In this paper, we discuss the UKIRT (United Kingdom Infra Red
Telescope) WFCAM (Wide Field CAMera) Transit Survey (WTS), the
first published wide-field near-infrared dedicated programme,
searching for short period ($<$10 day) transiting systems around M
dwarfs. The survey was designed to monitor a large sample ($\sim
10,000$) of low-mass stars with precise photometry. In
this paper we present an analysis of the first completed field in the
WTS. We demonstrate that we can already put useful constraints on the
hot Jupiter planet occurrence rate around M dwarf stars, and this is
currently the strictest constraint available.
\par The survey observing strategy is described in Section
\ref{sec:survey}. In Section \ref{sec:pipeline} a summary is given of
the data processing pipeline, describing the generation of final clean
lightcurves from raw exposures. In Section \ref{sec:mdwarfs} we
describe the M dwarf sample and discuss uncertainties in
classification. We evaluate the sensitivity of the survey in Section
\ref{sec:simulation} using transit injection and detection Monte Carlo
simulations. The lack of giant planets detected by the survey around M
dwarfs to date is discussed in Section \ref{sec:discussion}. Finally,
in Section \ref{sec:comparison}, we consider the
\citetalias{Howard_2012} sample and use it to place an upper limit on
the frequency of hot Jupiters around M dwarfs based on the Kepler Q2
data release. We discuss our WTS results in the context of both the
\cite{Bonfils_2011} and \citetalias{Howard_2012} studies.
\section{The WFCAM Transit Survey}
\label{sec:survey}
\par UKIRT is a 3.8m telescope, optimized for near-infrared
observations and operated in queue-scheduled mode. The Minimum
Schedulable Blocks are added to the queue to match the ambient
conditions (seeing, sky-brightness, sky transparency etc). The WTS
runs as a backup programme when observing conditions are not good
enough (e.g. seeing $>$ 1 arcsec) for main surveys such as UKIDSS
\citep{Lawrence_2007}.
\par The Wide Field Camera (WFCAM) comprises four Rockwell Hawaii-II
PACE arrays, with 2k$\times$2k pixels each covering 13.65 arcmin x
13.65 arcmin at a plate scale of 0.4 arcsec/pixel. The detectors are
placed in the four corners of a square with a separation of 12.83
arcmin between the chips. This pattern is called a {\em pawprint}.
%
\par
\begin{figure}
\begin{center}
\includegraphics[width=8cm]{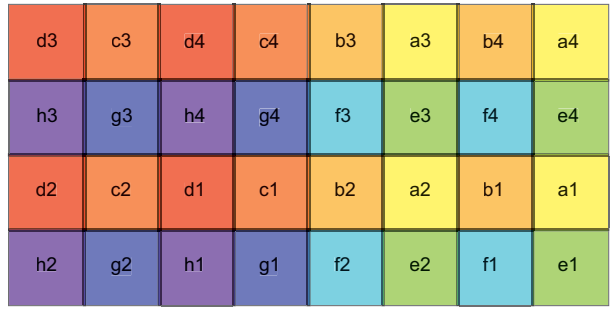}\\
\end{center}
\caption{Observation pattern for the WTS. A field of 1.6 square
  degrees consists of 8 pawprints (a-h), each pawprint is built up
  from the simultaneous exposures of the four (numbered)
  detectors. The X-axis is Right Ascension, and increases to the left
  of the figure, while the Y-axis is Declination, and increases to the
  top.}
\label{fig:pawprints}
\end{figure}

 The WTS time series data are obtained in the J band
($\lambda_\mathrm{eff}=1220\ \mathrm{nm}$), the fields were also
observed once in all other WFCAM bands (Z,Y,H,K) at the beginning of
the survey. This photometric system is described in
\cite{Hodgkin_2009}. Each field, covering 1.6 square degrees, is made
up of an 8 pawprint observation sequence with slightly overlapping
regions at the edges of the pawprints (Fig.\
\ref{fig:pawprints}). Observations are carried out using 10s exposures
in a jitter pattern of 9 pointings. A complete field takes 16 min to complete
forming the minimum cadence of the survey. Observing blocks typically
comprise 2 or 4 repeats of the same field. 
\par Four distinct WTS survey fields were chosen to be reasonably
close to the galactic plane to maximize stellar density while keeping
giant contamination and reddening at an acceptable level (see Section
\ref{sec:mdwarfs}). The four fields are distributed in right ascension
at 03, 07, 17 and 19 hours to provide all year coverage (at least one
field is usually visible). At the time of the analysis presented in
this paper, the 19 hour field has approximately 950 epochs which is
close to completion (the original proposal requested 1000 exposures
for each field), the other three are less complete.  A summary of the
key properties of the survey regions is shown in Table
\ref{tab:fields}.
\begin{table}
\centerline{
  \begin{tabular}{>{\small}c@{\hskip .5em}>{\small}c@{\hskip .5em}>{\small}c@{\hskip .5em}>{\small}c@{\hskip .5em}>{\small}c@{\hskip .5em}>{\small}c}
  \hline
  name & coordinates & galactic & No.\ of & objects & stellar \\
  & RA, DEC & l,b  & epochs & ($J<17$)  & objects \\
  & (h),(d)&(d),(d) & & & \\
\hline
19 & 19.58+36.44 & 70.03+07.83 & 950 & 69161 & 59270 \\
17 & 17.25+03.74 & 24.94+23.11 & 340 & 17103 & 15343 \\
07 & 07.09+12.94 & 202.89+08.91 & 350 & 24153 & 21224 \\
03 & 03.65+39.23 & 154.99-12.99 & 240 & 17221 & 15159 \\
%
\hline
\end{tabular}
}
\caption{Summary of the WTS fields and their coverage as of 27th May
  2010\label{tab:fields}. Stellar objects are
    morphologically identified by the photometric pipeline (see
    Sec.\ref{sec:catalogues}).}
\end{table}
\begin{figure}
\begin{center}
\includegraphics[width=9cm]{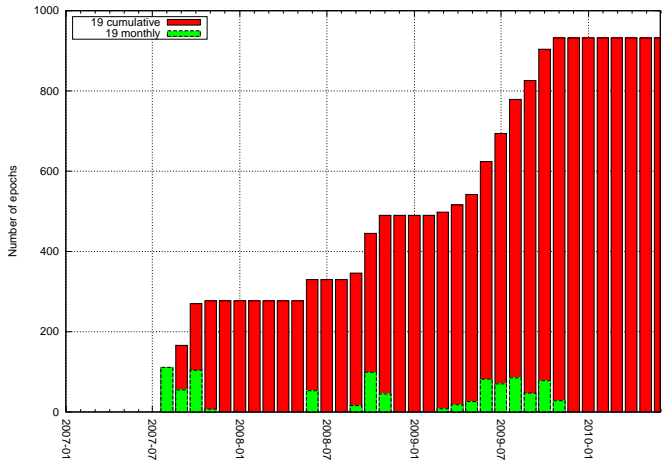}\\
\end{center}
\caption{Monthly and cumulative distributions of observational epochs
  in the 19hr field as of 27th May 2010. There are seasons 
  when observations were frequent with a handful of images taken every
  night and there are big gaps when the survey was not scheduled for
  observation.}
\label{fig:num_of_epochs}
\end{figure}
\par The WTS has a lower priority than most of the main UKIRT
programmes, thus observations are not distributed uniformly over
time. For any given field, such as the 19 hour field
(Fig.\ \ref{fig:num_of_epochs}), there are large gaps when the field
is not visible, as well as variations between and within seasons.
\section{Pipeline overview}
\label{sec:pipeline}
The WTS uses list-driven aperture photometry on processed images to
construct lightcurves from the stacked data frames at the pawprint
level. We describe this procedure briefly in this section.
\subsection{2D pipeline}
All images taken with WFCAM are processed using an image reduction
pipeline operated by the Cambridge Astronomical Survey Unit
(CASU)\footnote {\texttt
  {http://casu.ast.cam.ac.uk/surveys-projects/wfcam}}. The WFCAM
pipeline loosely evolved from strategies developed for optical
processing (e.g. the Wide Field Survey on the Isaac Newton Telescope,
\citep{Irwin_Lewis_2001}) and implements methods originally presented
in \cite{Irwin_1985}. The 2D processing is a fairly standardized
procedure for the majority of projects using WFCAM on UKIRT. We give a
brief overview of the image processing steps here.
\par Images are converted into multi extension FITS format, containing
the data of the 4 detectors in one pawprint as extensions. Other data
products from the pipeline (catalogues, lightcurves) are also stored
in binary FITS table files. A series of instrumental correction steps
is performed, accounting for: nonlinearities, reset-anomalies, dark
current, flatfielding (pixel-to-pixel), defringing and sky
subtraction. The sky subtraction removes any spatial variation in the
sky background but preserves its mean level. The sky background is
calculated as a robust $k\sigma$ clipped median for each bin in a
coarse grid of 64x64 pixels.\footnote{The median absolute deviation
  (MAD) is used as a robust estimator of the root mean square (RMS) in
  most pipeline components both during processing individual frames
  and lightcurves. For normal distribution,
  $\mathrm{RMS}=1.48\cdot\mathrm{MAD}$.} The sky background map is
filtered by 2D bilinear and median filters to avoid the sky level
shifting in bins dominated by bright objects. The final step is to
stack the 9 individual WTS exposures to produce one 90 second
exposure. Note that a simplified version of the catalogue generation
and astrometric calibration steps (described below) are run on the
individual unstacked exposures to ensure that they can be aligned
before combining.
\subsection{Catalogues, Astrometry and Photometry}
\label{sec:catalogues}
Object detection, astrometry, photometry and classification are
performed for each frame. Object detection follows methods outlined in
\cite{Irwin_1985} (see also
\cite{Lawrence_2007}). Background-subtracted object fluxes are
measured within a series of soft-edged apertures (i.e.\ pro-rata
division of counts at pixels divided by the aperture edge). In the
sequence of apertures, the area is doubled in each step. The scale
size for these apertures is selected by defining a scale radius fixed
at 1.0 arcsec for WFCAM. A 1.0 arcsec radius is equivalent to 2.5
pixels for WTS non-interleaved data. In 1 arcsec seeing an
rcore-radius aperture contains roughly 2/3 of the total flux of
stellar images. Morphological object classification and derived
aperture corrections are based on analysis of the curve of growth of
the object flux in the series of apertures \citep{Irwin_2004}.
\par Astrometric and photometric calibrations are based on matching a set
of catalogued objects with the 2MASS \citep{2MASS} point source
catalogue for every stacked pawprint. The astrometry of data frames
are described by a cubic radial distortion factor (zenithal polynomial
transformation) and a six coefficient linear transformation allowing
for scale, rotation, shear and coordinate offset corrections. Header
keywords in FITS files follow the system presented in
\cite{Greisen_2002,Calabretta_2002}. The photometric calibration of
WFCAM data is described in \cite{Hodgkin_2009}. These calibrations
result in the addition of keywords to the catalogue and image headers,
enabling the preservation of the data as counts in original pixels and
apertures.
\subsection{Master catalogues}
\par Following the standard 2D image reduction procedures, catalogue
generation and calibration, we have developed our own WTS lightcurve
generation pipeline. This is largely based on previous work for the
Monitor project described in \cite{Irwin_2007} where more technical
details are given.
\par As a first step in the lightcurve pipeline, master images are
created for each pawprint by stacking the 20 best-seeing photometric
frames.  The master images play dual roles in our processing: they
define the catalogue of objects of the survey for each pawprint with
fixed coordinates and identifications numbers (IDs). Thus the
source-IDs will never change for the WTS, however their coordinates
could if we were to include a description of their proper
motions. This has not yet been done.
\par Source detection and flux measurement is performed for each
master image to create a series of master catalogues, and astrometric
and photometric calibration recomputed, again with respect to
2MASS. These object positions are then fixed for the survey. Each
source has significantly better signal-to-noise on the master image,
and thus better astrometry (from reduced centroiding errors) than
could be achieved in a single exposure. As discussed in
\cite{Irwin_2007}, centroiding errors in the placement of apertures
can be a significant source of error in aperture photometry,
particularly for undersampled and/or faint sources.
\par This master catalogue is then used as an input list for
flux-measurement on all the individual epochs, a technique we call
list-driven photometry.  A WCS transformation is computed between the
master catalogue and each individual image using the WCS solutions
stored in the FITS headers. Any residual errors in placing the
apertures will typically be small systematic mapping errors that
affect all stars in the same way or vary smoothly across frames. In
practice, this can be corrected by the normalization procedure (see
Section \ref{sec:normalization}). The same soft-edged apertures are
used (as described above), except that the position of the source is
no longer a free parameter. Thus for each epoch of observation, a
series of fluxes for the same sources is measured.
\subsection{Lightcurve Construction and Normalization}
\label{sec:normalization}
\par Although the photometry of each frame is calibrated individually
to 2MASS sources \citep{Hodgkin_2009}, these values can be refined for
better photometric accuracy. Lightcurves constructed using the default
calibrations typically have an $RMS$ at the few percent level (for
bright unsaturated stars). To improve upon this, an iterative
normalization algorithm is used to correct for median magnitude
offsets between frames, but also allowing for a smooth spatial
variation in those offsets. In each iteration, lightcurves are
constructed for all stellar objects and a set of bright stars selected
($13<J<17$) excluding the most variable decile of the group based on
their (actual iteration) lightcurve RMS. Then for each frame, a
polynomial fit is performed on the magnitude differences between the
given frame magnitudes and the corresponding median (lightcurve)
magnitude for the selected objects. The polynomial order is kept at 0
(constant) until the last iteration. In the last iteration, a second
order, 2D polynomial is fitted as a function of image coordinates. For
each frame, the best fit polynomial magnitude correction is applied
for all objects and the loop starts again until there is no further
improvement. Multiple iterations of the constant correction step help
to separate inherently variable objects from non-variable ones
initially hidden by lightcurve scatter caused by outlier frames. The
smooth spatial component during the last iteration accounts for the
effects of differential extinction, as well as possible residuals from
variation in the point spread function (PSF) across the field of view.
\par It is also found that lightcurve variations correlate with
seeing. In an additional post-processing step, for each object, a second
order polynomial is fitted to differences from median magnitude as a
function of measured seeing. Magnitude values are then corrected by
this function on a per-star basis.
\subsection{Bad epoch filtering}
\par
\begin{figure*}
\centerline{\includegraphics[width=6cm]{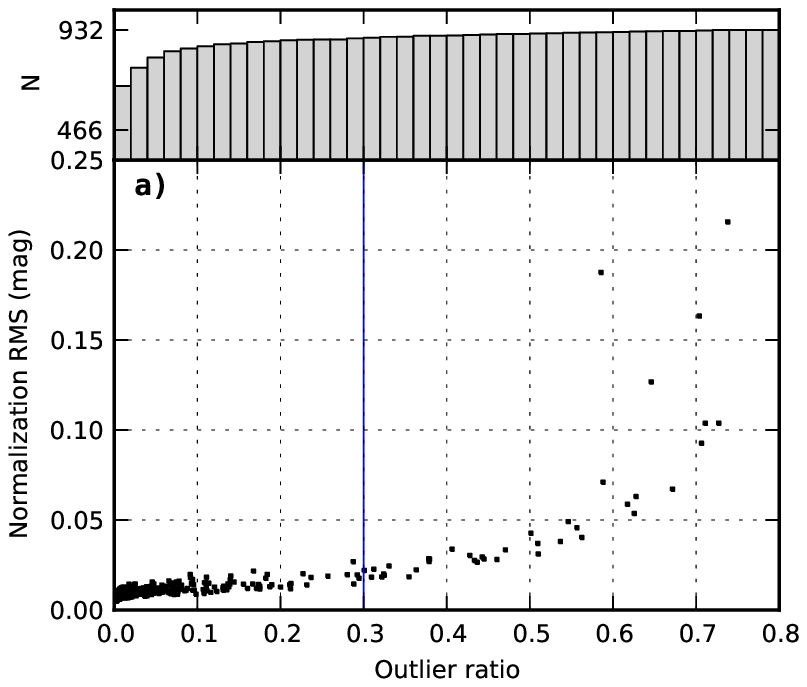}
\includegraphics[width=6cm]{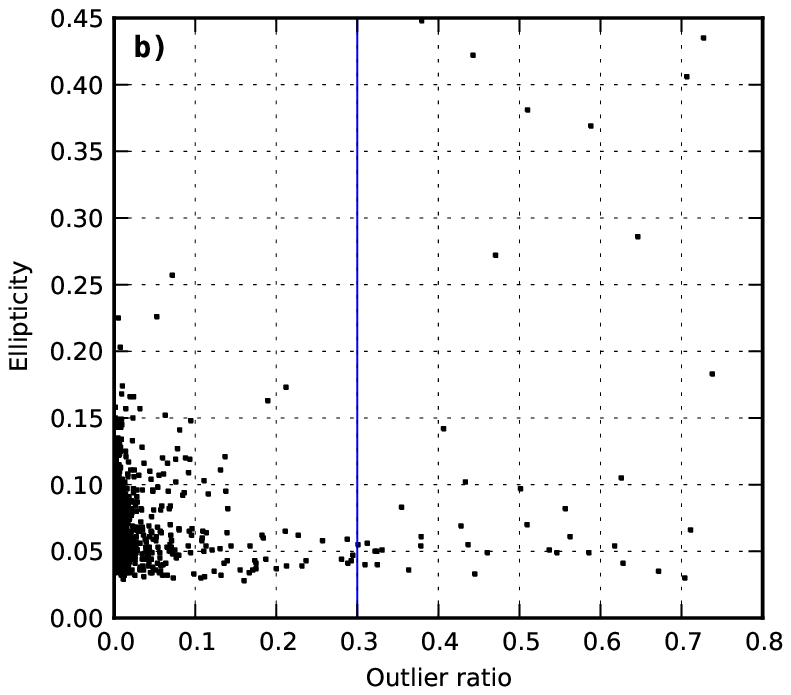}
\includegraphics[width=6cm]{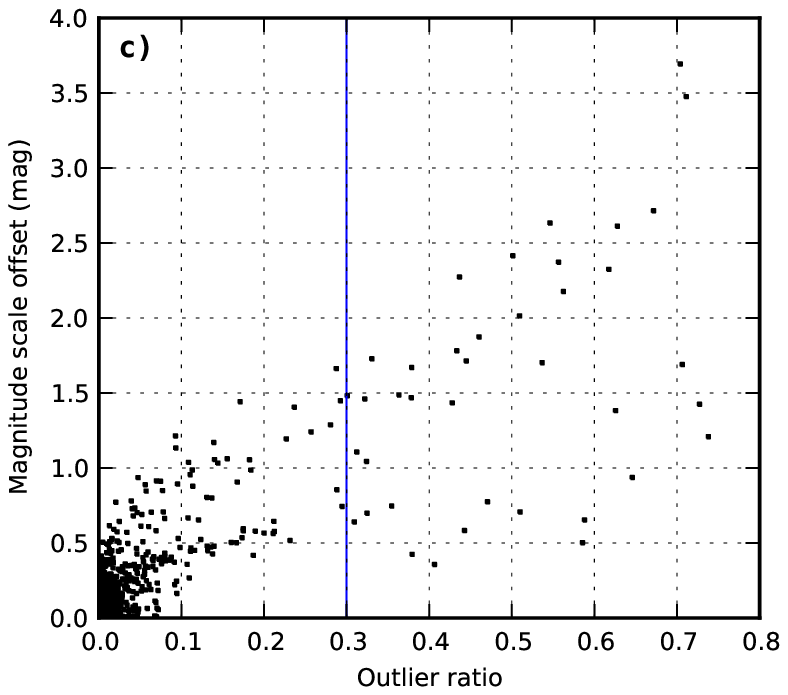}}
\caption{Per frame residual normalization RMS (a), average stellar
  ellipticity (b) and overall magnitude scale correction (c) as a
  function of $3\sigma$ outlier object ratio in the 19a
  pawprint. Epochs above a threshold of 0.3 are removed from the
  survey's lightcurve release and candidate search. See text for more
  details.}
\label{fig:outlier_frames}
\end{figure*}

 Data are taken in a wide range of observing conditions, sometimes
with bad seeing ($\sim 2$ arcseconds or worse) or significant cloud
cover. Additionally some frames are affected by loss and recovery of
guiding or tip-tilt correction during the exposure. We identify and
remove bad observational epochs that add outlier data points for a
significant number of objects in any chip of a pawprint. Where a
single (corrected) epoch has in excess of 30 per cent of objects
deviating by more than $3\sigma$ from the median flux we flag and
remove this epoch from all lightcurves. Fig.\ \ref{fig:outlier_frames}
shows some of the per-epoch parameters (which we store in the
lightcurve files) as a function of $3\sigma$ outlier object ratio. The
number of outlying objects has a strong correlation with the residual
RMS of the second order polynomial normalization (panel a) which is a
measure of the photometric unevenness of the image. The (cumulative)
frame magnitude offset applied during the normalization (panel c)
shows two distinct branches, and a large scatter can be seen in the
average frame ellipticity (panel b). These two panels help to identify
the main causes of bad epochs. High ellipticities arise from frames
with tracking/slewing problems, while high magnitude corrections are
suggestive of thick (and probably patchy) cloud. Our rejection
threshold is a compromise between the number of affected frames and
frame quality. At the outlier ratio threshold of 0.3, 39 frames (out
of 950, or 4 per cent) are removed in the 19hr field.
\subsection{Lightcurve quality}
\par
\begin{figure*}
\centerline{
\includegraphics[width=9cm]{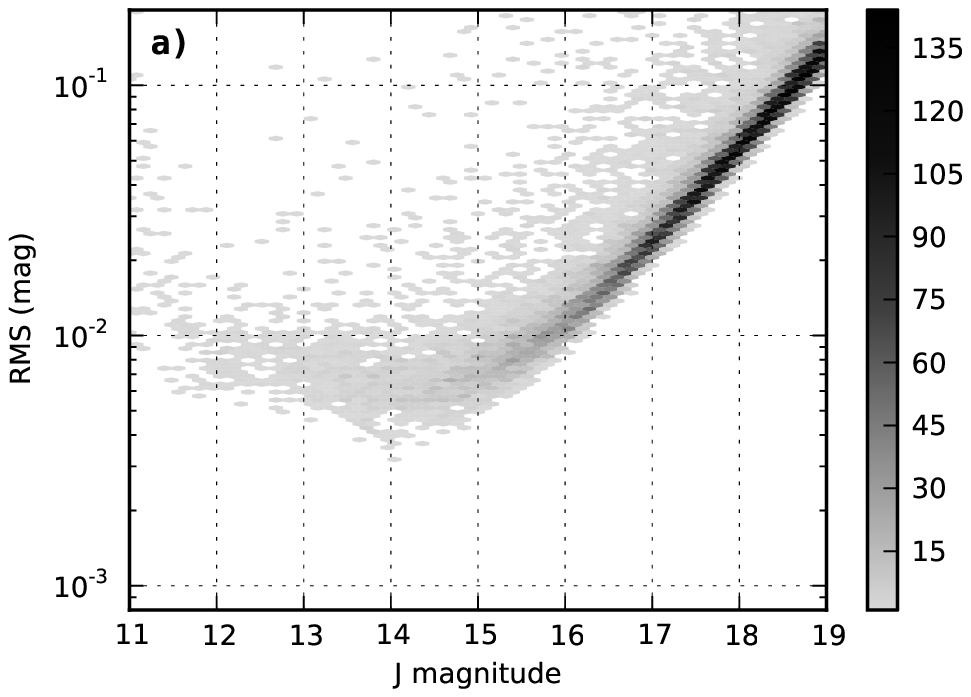}
\includegraphics[width=9cm]{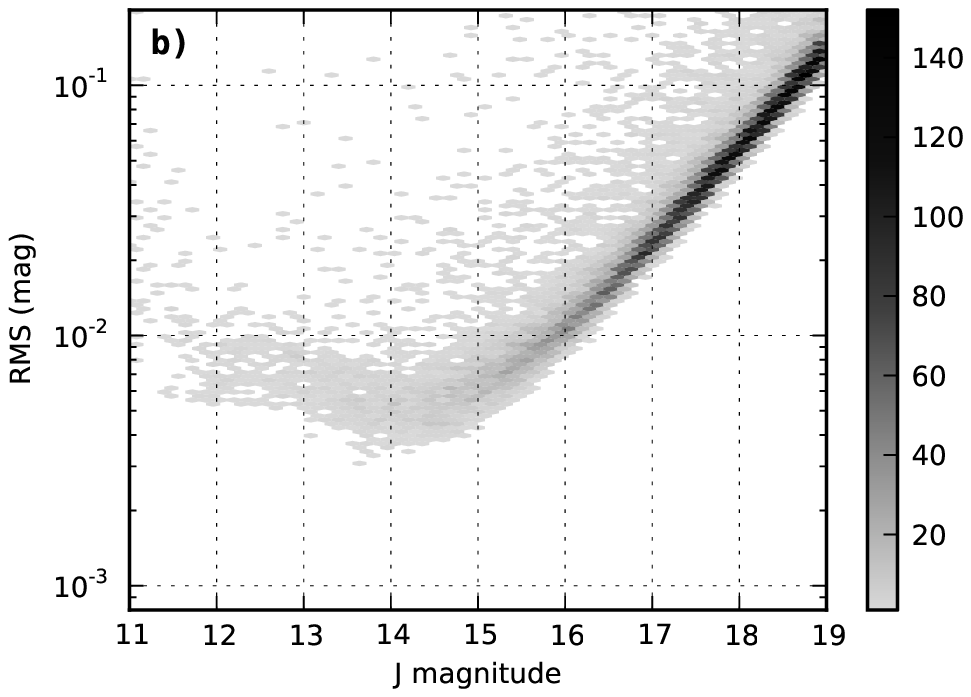}}
\centerline{\includegraphics[width=9cm]{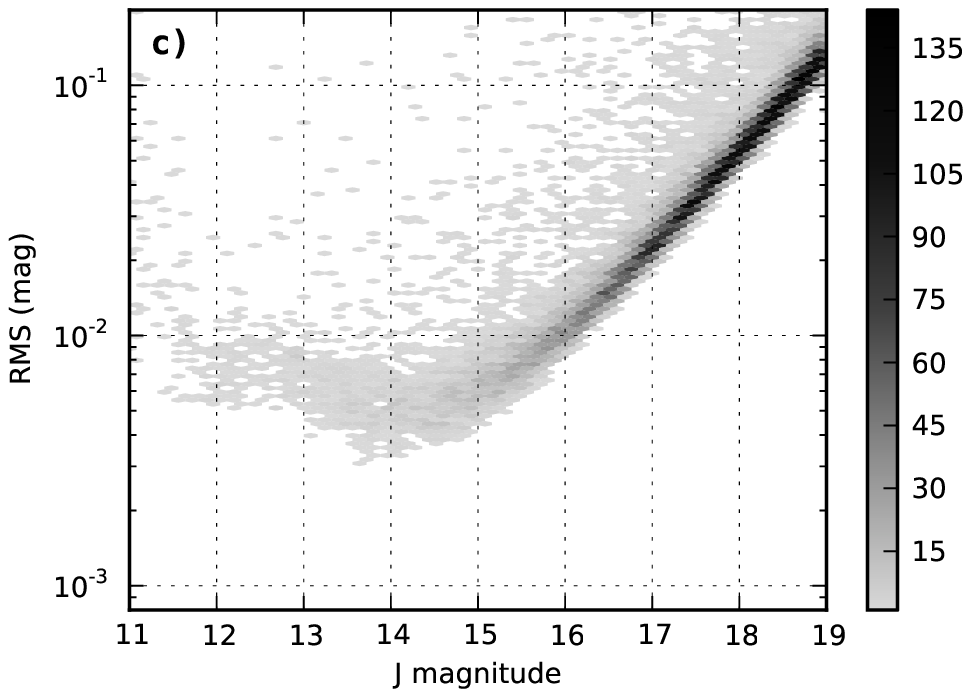}
\includegraphics[width=9cm]{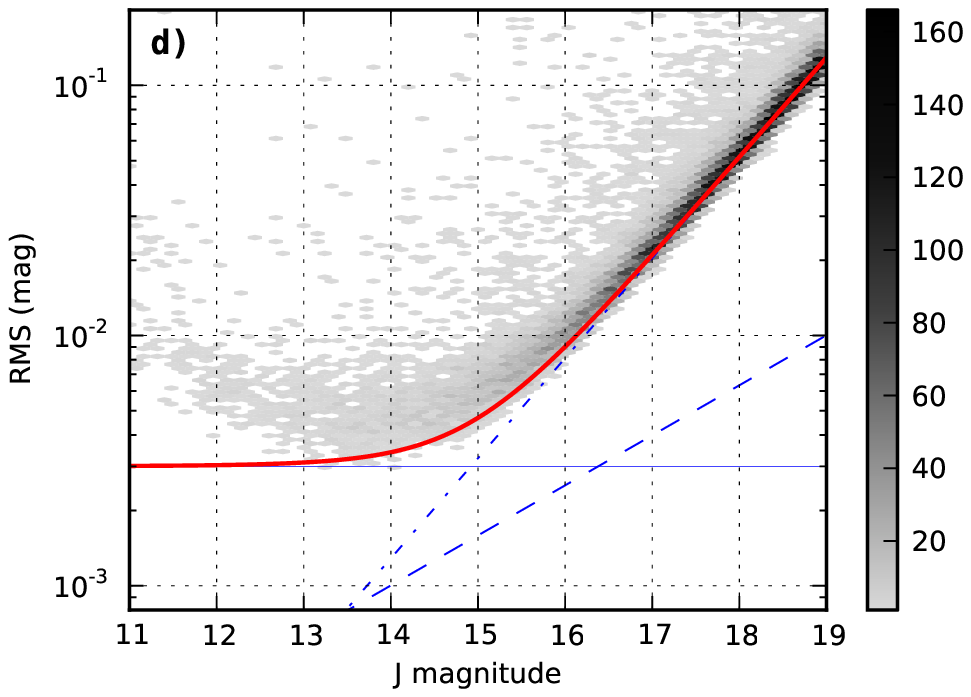}
}
\caption{RMS of stellar objects in the 19a pawprint with different
  pipeline optimizations; a) constant normalization b) quadratic
  normalization c) outlier frame filtering d) seeing correction. In
  panel d) a noise model (thick red solid line) consisting of Poisson noise
  (dashed line), sky noise (dash-dotted line) and systematic noise of
  3 mmag (thin blue solid line) is drawn.}
\label{fig:rms_pipeline}
\end{figure*}

 In Fig.\ \ref{fig:rms_pipeline} and Table
\ref{tab:pipeline_steps} we summarize the WTS lightcurve quality at
different pipeline steps. In each step we change or add one pipeline
feature. The normalization procedure of the photometric scale by
per-frame constant offsets (a), the quadratic spatial correction (b)
during the normalization, the bad epoch filtering (c) and the seeing
correction (d) give improvements at the several mmag level for
unsaturated bright stars. In panel
(d) a theoretical noise model curve consisting of Poisson noise, sky
noise and a constant residual are drawn. 
A  constant systematic error of 3 mmag is applied to the model shown
in Fig.\ \ref{fig:rms_pipeline} to bring the model roughly into
agreement with the data. This should be seen as the minimum systematic
error in our lightcurves. Some saturation appears and makes lightcurve
RMS worse for objects brighter than $J=13$ while for the faint end,
the sky noise dominates. 
\begin{table*}
\begin{center}
\begin{tabular}{ccccccc}
\hline
$J=$  & 11-12 & 12-13 & 13-14 & 14-15 & 15-16 & 16-17 \\
\hline
(a) constant normalization & 9.3 & 6.1 & 5.7 & 6.6 & 9.7 & 19.1 \\
(b) quadratic normalization & 9.1 & 5.8 & 5.0 & 5.9 & 9.2
& 18.8 \\
(c) + outlier filtering & 8.7 & 5.6 & 4.9 & 5.7 & 8.8 & 18.1 \\
(d) + seeing correction & 7.3 & 4.8 & 4.6 & 5.6 & 8.5 & 17.6 \\
\hline
\end{tabular}
\end{center}
\caption{Median lightcurve RMS (mmag)
  as a function of object magnitudes using different pipeline options
  for the 19a pawprint. See text for details, and
  Fig.\ \ref{fig:rms_pipeline} for illustration.}
\label{tab:pipeline_steps}
\end{table*}
\subsection{Transit detection}
\label{sec:transdet}
\par To detect transit signals, we use a variant of the boxcar-fitting
(Box Least Squares, or BLS) algorithm developed by
\cite{Aigrain_Irwin_2004}. They derive 
the transit fitting algorithm starting from a maximum likelihood
approach of fitting generalized periodic step functions to the
lightcurve. It was demonstrated that for planetary transits a simple
box shaped function is sufficient. The algorithm in this form is
equivalent to BLS developed by \cite{Kovacs_2002}. The signal to red
noise\footnote{Noise that includes both uncorrelated (white) and
  correlated (red) components. In some cases, this is called `pink'
  noise in the literature.}  statistic is used to measure transit
fitting significance (eq.\ 4 reproduced from \cite{Pont_2006}):
\begin{equation}
S_\mathrm{red}=\frac{d}{\sqrt{\frac{\sigma_0}{n}+\frac{1}{n^2}\sum_{i\not=j}C_{ij}}}
\end{equation}
where $n$ is the number of in-transit data points in the whole
lightcurve, $\sigma_0$ is the measurement error of the individual data
points, $d$ is the fitted depth of the BLS algorithm, $C_{ij}$ is the
covariance of two in-transit data points. Also following
\cite{Pont_2006} we adopt a detection threshold of
$S_\mathrm{red}=6$. Objects that pass this threshold are considered
candidate transiting systems.
%
%
\section{A sample of M dwarfs}
\label{sec:mdwarfs}
\subsection{Identification of M dwarfs}
\par The high numbers of objects in the WTS and their faint magnitudes
make spectral classification of all WTS objects potentially resource
consuming. Instead, we use homogeneous broadband optical (from SDSS
DR7, \cite{SDSS_DR7}) and near-infrared (WFCAM) photometry to estimate
reliable stellar spectral types. Specifically, effective temperatures
are measured from fitting NextGen stellar evolution models
\citep{Baraffe_1998} to $g$$r$$i$$z$ (SDSS) and $Z$$Y$$J$$H$$K$
(WFCAM) magnitudes. For comparison, we also fit the Dartmouth
\citep{Dotter_2008} models, but in this case we use 7 passbands (Z and
Y are not available). Least squares minimization is performed for the
available photometry, fitting for temperature and a constant magnitude
offset (distance modulus) as model parameters. The model grid data is
smoothed by a cubic interpolation. We follow the
temperature--spectral-class relation in Table 1 of
\cite{Baraffe_1996}: i.e. 3800K, 3400K, 2960K and 1800K corresponding
to spectral types of M0, M2, M4 and M9 respectively.
%
\begin{figure*}
\centerline{\includegraphics[width=9cm]{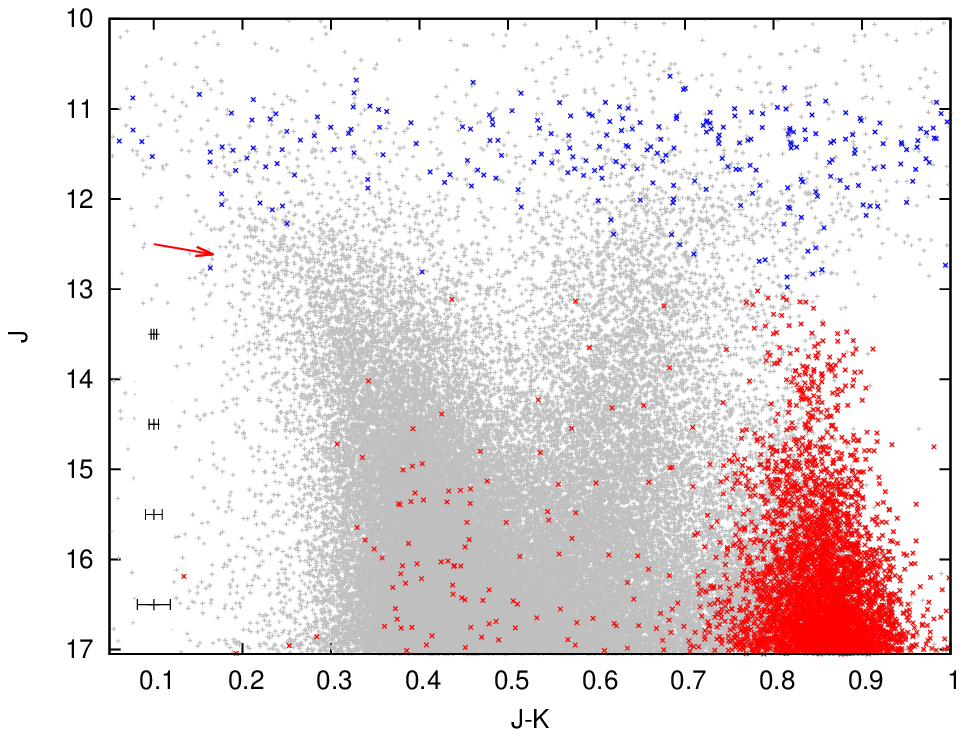}
\includegraphics[width=9cm]{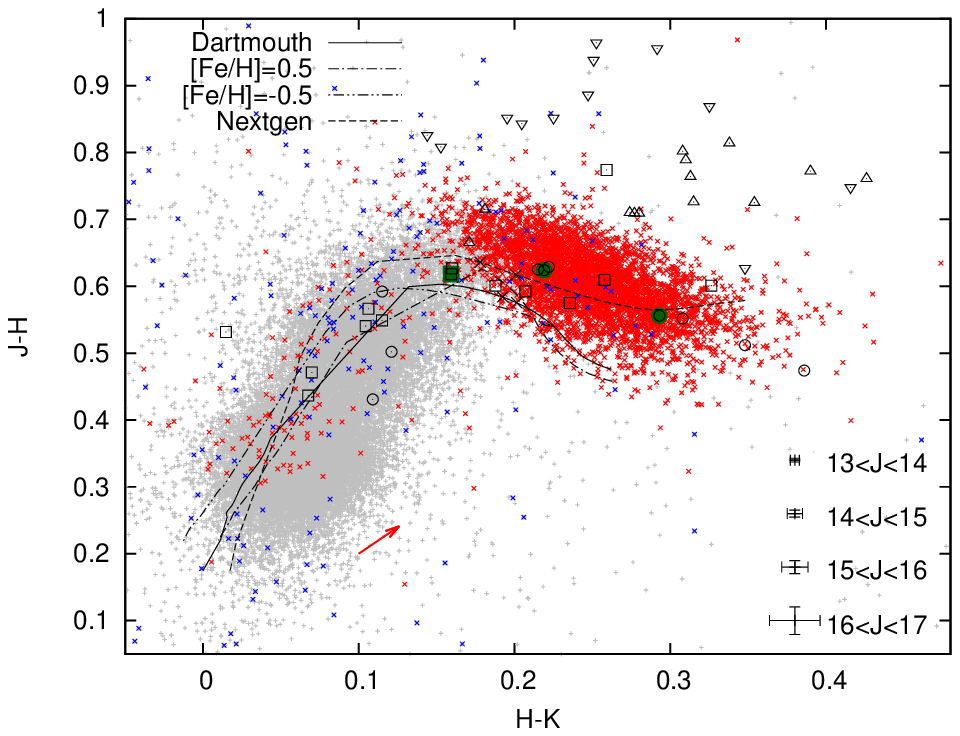}}
\centerline{\includegraphics[width=9cm]{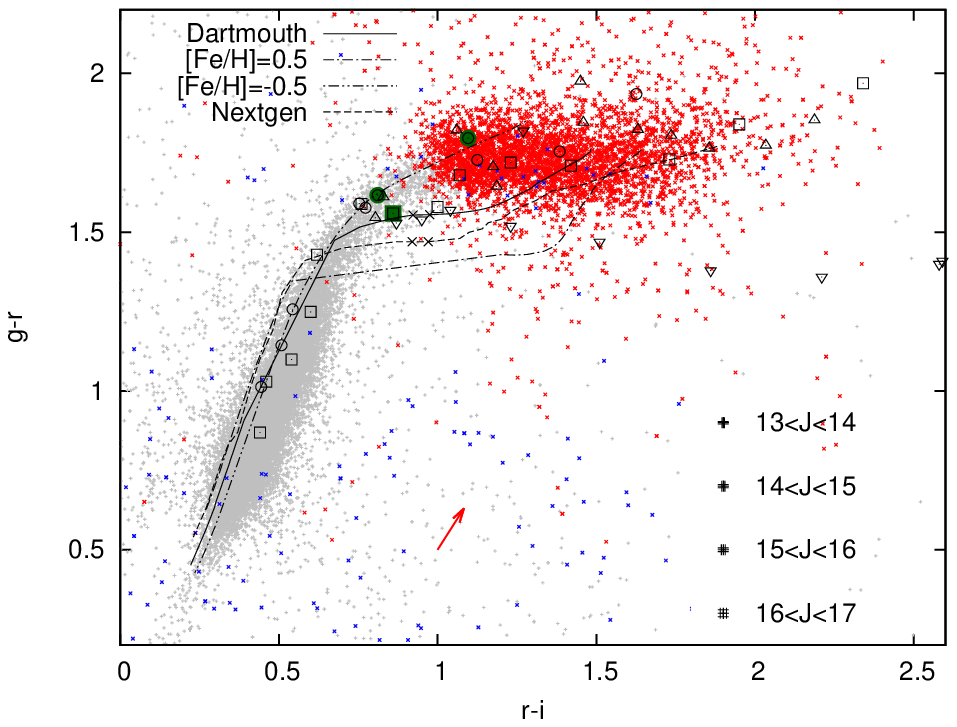}
\includegraphics[width=9cm]{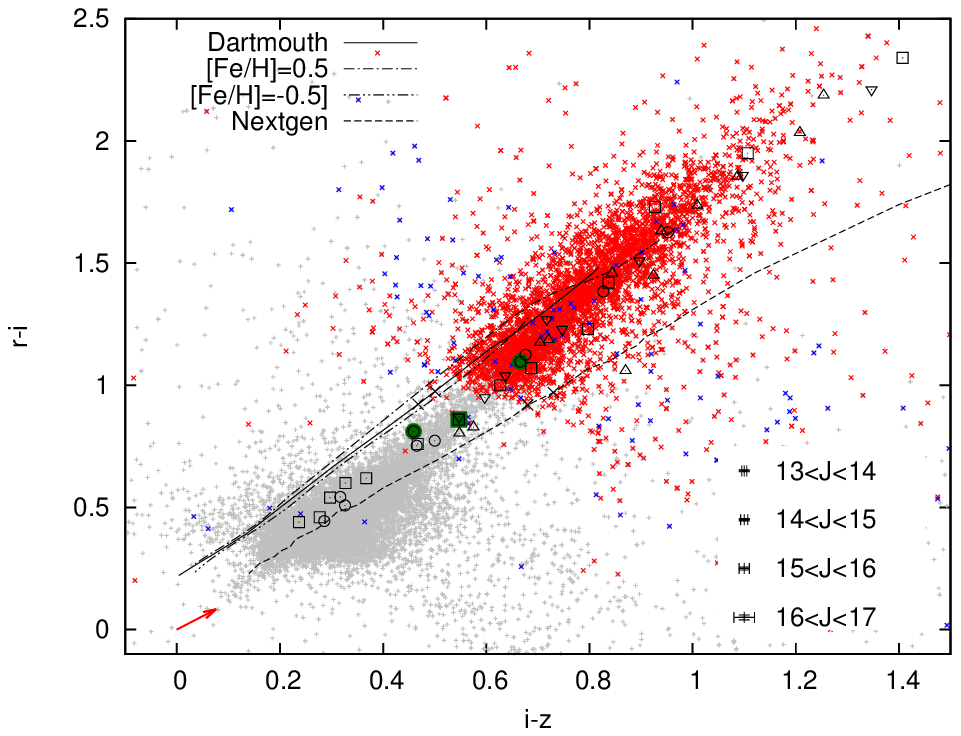}}
\caption{Colour-magnitude and colour-colour plots of the M dwarf
  sample ($13<J<17$, red crosses; $11<J<13$, blue crosses) of this
  study and other stellar objects ($10<J<16$) in the 19hr field based
  on WFCAM and SDSS data. A sequence of spectroscopic K and M dwarfs
  are shown from \protect\cite{Covey_2007} ($\square$) and
  \protect\cite{Hewett_2006} ($\circ$), respectively. Green markers
  show the M0 dwarf members of these sequences. Triangles
  ($\bigtriangledown$,$\bigtriangleup$) are M giants (III) from the
  same studies. Estimated maximum interstellar extinctions in
  different colours are marked by red arrows. Error crosses are from
  the corresponding catalogues.}
\label{fig:mdwarfs_cc_plot}
\end{figure*}
\par Fig.\ \ref{fig:mdwarfs_cc_plot} shows WFCAM and SDSS
colour-magnitude and colour-colour panels of the stellar objects in
the 19hr field. The M dwarf sample identified by the SED fitting on
NextGen magnitudes is marked by red ($13<J<17$) and blue ($11<J<13$)
crosses. A small number (6 per cent) of objects have outlying colour
values due to saturation in one or more WFCAM or SDSS filters
(typically $J<13$). These magnitudes are sigma-clipped during the SED
fitting procedure. The panels also show 1 Gyr isochrones from the
NextGen and Dartmouth models in the 2000K--6500K and 3200K--7700K
temperature intervals respectively. The solid and dashed curves
correspond to solar metallicity, the dash-dot to a metal-rich
([Fe/H]=+0.5), the dash-double-dot to a metal-poor ([Fe/H]=-0.5) model
isochrone of the Dartmouth model respectively.  Temperatures of 3800K
and 3900K are marked (x) on the isochrones.  Comparison between
Dartmouth model isochrones show little significant difference in model
colours between ages of 250Myr and 5Gyr. For very young stars, we
might expect to pick up significant colour variation, however we
expect very few very young (age $< 100$ Myr) stars in our survey
field. In fact \cite{Ciardi_2011} analyse the very nearby Kepler field
and find that the low-mass dwarf population is dominated by young thin
disk stars, thus our selected age of 1Gyr is reasonable.
\par The figure also presents measured colours of K and M dwarfs and
MIII giants of the Pickles photometric standards (dwarfs: $\square$,
giants: $\bigtriangledown$) from \cite{Covey_2007} and of the
Bruzual-Persson-Gunn-Stryker atlas (dwarfs:$\circ$,
giants:$\bigtriangleup$) from \cite{Hewett_2006}. All panels show
colours in the Vega system, AB-Vega offsets are taken from Table 7 in
\cite{Hewett_2006}, 2MASS-WFCAM conversions are calculated by
relations given by \cite{Hodgkin_2009}. We denoted by filled green
markers the M0 dwarf members (one and two objects, respectively) of
these observations. Our identified M dwarfs are separated well from
the sample of M giants ($\bigtriangledown$,$\bigtriangleup$) in the
J-H vs.\ H-K and g-r vs.\ r-i panels, so we expect a low giant
contamination level in our sample.  The panels in Fig.\
\ref{fig:mdwarfs_cc_plot} also demonstrate some difficulties in
identifying M dwarfs. Model predictions do not reproduce observed
patterns in all colour combinations. The NextGen isochrone has the
best agreement in the infrared (J-H vs.\ H-K) while the Dartmouth
models fit the r-i vs.\ i-z colours rather better. We note that the
Nextgen colour predictions are too blue in the optical bands for low
temperatures, which is a known model attribute \citep[e.g. by 0.5 mag
in V-I, ][]{Baraffe_1998}.
\par We also note that based on the residual $\chi^2$
  values during the SED fitting procedure the photometric errors from
  the catalogues (shown in the colour-colour panels of
  Fig.\ref{fig:mdwarfs_cc_plot}) were found to be underestimated. We
  assumed a systematic error between the WFCAM and SDSS catalogues and
  added a 0.03 mag systematic term in quadrature to the individual
  magnitude errors (used as weights in the fitting).
\par The Dartmouth data sets cover a narrower temperature range with a
lowest temperature of 3200K. This makes the dataset unsuitable for
selecting later (M4 and later) M dwarfs, although it gives a useful
comparison for warmer stars. We conclude that for early M dwarfs
(M0-2), the Dartmouth and Nextgen models give rise to very similar
selections assuming solar metallicity. In the metal poor and metal
rich cases the Dartmouth isochrones give about 40 per cent increase
and 30 per cent decrease in numbers of early M dwarfs respectively.
\par In the g-r vs. \ r-i diagram, we note that the scatter in g-r for
objects with r-i$>1$ is larger (by about 0.2 mags at g=20) than can be
explained solely from the SDSS photometric errors. This enhanced
scatter can be explained by reddening alone (see below), and needs no
significant spread in metallicity.
\par \cite{Leggett_1992} analyzed photometry of 322 M dwarfs and found
that the effects of metallicity can be seen in M dwarf infrared colours (I-J,
J-K, I-K, J-H, H-K) while not discernible in visible colours (U-B,
V-I, B-V). They also noted that this feature is not reproduced by
evolutionary models of \cite{Mould_1976,Allard_1990}. We 
note that the Dartmouth model also shows a much smaller effect than
seen in \cite{Leggett_1992} in the infrared, but a very large effect
in g-r vs r-i.
\par In our paper, we base object classification on the NextGen
model. Out of the 59000 morphologically identified stars in the
$11<J<17$ magnitude range, we identified $\sim 4600$ M dwarfs. A more
in-depth modelling of the objects in the WTS is an ongoing effort.
\subsection{Interstellar extinction}
\par During the SED fitting procedure, we cannot fit
  for the interstellar reddening and for the distance modulus on a per
  object basis due to being the fit badly determined. Therefore, we
  make an upper estimation for the reddening in our sample. We estimate
a maximum distance for target M dwarfs of 1.5 kpc by assuming that one
of our intrinsically brightest objects (an M0, M$_{\rm J}=6$ from
NextGen) is observed at the faint limit of the study ($J=17$). $A_V$
extinction values are calculated from the Galactic model of
\cite{Drimmel_2003}. The three dimensional Galactic model consists of
a dust disk, spiral arms mapped by HII regions and a local
Orion-Cygnus arm segment. They use COBE/DIRBE far infrared
observations to constrain the dust parameters in the model. We use the
code provided to determine $A_V$ at 1.5 kpc. To calculate absorption
in UKIRT and SDSS bandpasses, we use conversion factors $A/A_V$ from
Table 6 of \cite{Schlegel_1998} that evaluates the reddening law of
\cite{Cardelli_1989} and of \cite{ODonnell_1994} for the infrared and
the optical bands respectively. The reddening laws assume $R_V\equiv
A_V/E_{B-V}=3.1$, an average value for the diffuse interstellar medium
\citep{Cardelli_1989}. Red arrows show our reddening estimates in
different colours in Fig.\ \ref{fig:mdwarfs_cc_plot} ($A_V:$ 0.409,
$A_J$: 0.113, $E_{J-K}$: 0.067, $E_{J-H}$: 0.041, $E_{H-K}$: 0.026,
$E_{g-r}$: 0.130, $E_{r-i}$: 0.083, $E_{i-z}$: 0.076). The total
Galactic extinction in the 19hr field direction from
\cite{Schlegel_1998} is $A_V= 0.439$\footnote{Provided by the Nasa
  Extragalactic Database. \texttt{http://ned.ipac.caltech.edu/}}.
%
%
%

%
\subsection{M dwarf bins}
%
%
%
\label{sec:mdwarf_bins}
\par Stellar radius changes significantly between early- and late-type
M dwarfs. For our sensitivity simulation purposes, we can use a linear
approximation for the mass-radius relationship for M dwarfs:
$M_{*}/M_\mathrm{\odot} \approx R_{*}/R_\mathrm{\odot}$. This slightly
deviates from the NextGen mass-radius prediction \citep{Baraffe_1998}
but it is in good agreement with observed M dwarf radii in
\cite{Kraus_2011}.  It is important in our simulations to treat early
M dwarfs distinctly from later smaller M dwarfs. Therefore we divided
our M dwarf sample into three coarse spectral bins. In the
19hr field, 2844 stars were identified as M0-2 ($3800\mathrm{K} >
T_\mathrm{eff} > 3400\mathrm{K}$), 1679 as M2-4 ($3400\mathrm{K} >
T_\mathrm{eff} > 2960\mathrm{K}$) and 104 as M4-9 ($2960\mathrm{K} >
T_\mathrm{eff} > 1800\mathrm{K}$). The third, latest (M4-9) type bin
is very sparsely populated (2 per cent of the M dwarf 
sample) and numbers suffer from high uncertainty, so this bin will be
omitted from the simulations, leaving us with two subsamples.  
\par Considering the `colour distance' along the isochrones between
the boundary of our identified object groups, the Pickles M0 members
(green squares) and the dependency of the isochrones on model
parameters, we estimate the bin temperature edges to be uncertain
about 250K.
\subsection{M Dwarf Lightcurves}
\label{sec:mdwarf_rms}
\par
\begin{figure*}
\centerline{\includegraphics[width=\textwidth]{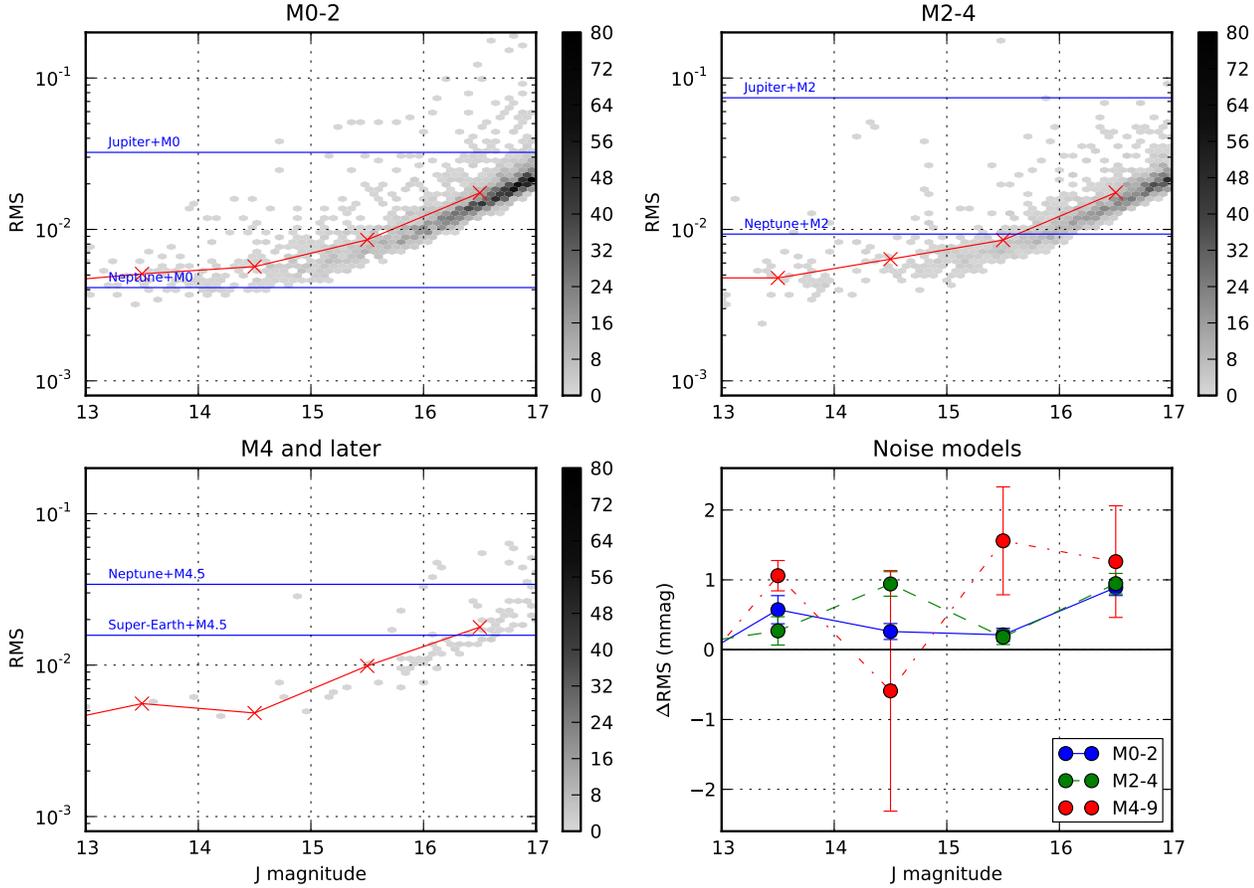}}
\caption{Comparison of lightcurve RMS in the M dwarf spectral
  groups. The solid curve shows median values of the RMS per magnitude
  bin. The bottom right panel shows the differences of
    the median RMS model curves from the median RMS of the earlier
    type stars. (M0-2: solid, M2-4: dashed, M4-9: dash
    dotted). Signal depths of edge-on transiting systems with
  Neptune, Jupiter or Super-Earth ($2.7R_\oplus$) size planets are
  marked.}
\label{fig:rms_mtypes}
\end{figure*}

 M dwarfs are known to be intrinsically more variable than more
massive main sequence stars primarily due to spots \citep[e.g.\
][]{Chabrier_2007,Ciardi_2011}. The fraction of active M dwarfs also
rises towards later (M4-9) subtypes \citep{West_2011}. We examine the
variability of the M dwarfs compared to the rest of the stellar sample
in Table \ref{tab:rms} and Fig.\ \ref{fig:rms_mtypes}.  The
photometric precision is at the 4-5 mmag level for brighter objects
($J<15$) and at the percent level for the fainter region ($15< J<17$,
86 per cent of the M dwarfs). We compare the binned median RMS values
of the three M subtypes to those of the warm (T$>$3800K)
stars. Although the effect is small, we do find evidence that the
median RMS for all three M dwarf samples is systematically higher,
although only at about the 0.5-1 mmag level, at all magnitudes (see
Fig.\ref{fig:rms_mtypes}).
\par We also calculate the fraction of $3\sigma$ outliers (a proxy for
the most variable objects) for the different subsamples.  We find that all three
M dwarf samples, as well as the warmer stars exhibit a ten per cent
outlier fraction.
\par Finally, these findings are also supported by 2D
Kolmogorov-Smirnov tests \citep{KS_2D}. The RMS versus magnitude
distributions are found to be consistently different for each of the M
dwarf samples versus the warm stars to very high confidence (test p
values $<10^{-3}$). These results add support to the argument that
observing in the J band gives limited sensitivity to spots
\citep{Goulding_2011}.
\par Typical signal depths for edge-on planetary systems are also
shown in Fig.\ \ref{fig:rms_mtypes} to give a rough impression of our
sensitivity. We expect to be able to detect Jupiter size planets in
edge-on systems around all of our M dwarfs (in all spectral bins:
M0-2, M2-4, M4-9), and Neptunes only in the M2-4 and M4-9 bins. As
stated above, the latest type bin (M4-9) is sparsely populated, and
omitted from the remaining analysis.
\begin{table}
\begin{tabular}{cccccc}
\hline
 $J=$     & 12-13 & 13-14 & 14-15 & 15-16 & 16-17 \\
\hline
M0-2 & 4.4 & 5.1 &5.7  & 8.5 & 17.4 \\
M2-4 & 4.8 & 4.8 & 6.4 & 8.5  &17.5 \\
M4-9 & 3.9 & 5.6 & 4.8 & 9.9  &17.8 \\
Earlier types & 4.8 & 4.5 & 5.4 & 8.3 & 16.6 \\
\hline
\end{tabular}
\caption{Median RMS (mmag) of M dwarf lightcurves and of earlier
  dwarfs for different object magnitudes.\label{tab:rms} }
\end{table}
\subsection{The WTS sample compared to other surveys}
The WTS is not the only transit survey specifically aimed at
discovering planets around M dwarfs, and it is useful to place our
survey in context. The MEarth survey individually targets about 2000
bright M dwarfs in a custom $i+z$-band filter \citep{Berta_2012}. The
survey was designed to focus on the nearest and brightest mid M 
dwarfs and has discovered one Super Earth size planet at the time of
writing \citep{Charbonneau_2009}.
\par The Palomar Transient Factory (PTF, \cite{Law_2012}) is a project
targeting M dwarfs in the R band. Their goal is to observe a total of
100,000 M dwarfs during the survey lifetime collecting about 300
epochs in each observing period. The PTF operates in roughly the same
magnitude range (specifically over R=14--20) as the WTS and has a
similar, seasonal observing schedule.
\par The Pan-Planets project also focuses on late-type stars. Their
goal is to target approximately 100,000 M dwarfs in 40 square
degrees. The programme collected about 100 hours of observations using
the Panoramic Survey Telescope and Rapid Response System (PanSTARRS)
so far \citep{Koppenhoefer_2009}.
\par The Next-Generation Transit Survey (NGTS) is also an upcoming
initiative to target a high number of bright M dwarfs ($V$<13) using
wide field cameras. The prototype instrument observed $\sim$ 400 early
and $\sim$ 50 late M dwarfs in 8 square degrees FOV at the one per cent
photometric precision level \citep{Chazelas_2012}.
\par Regarding space missions, the Kepler Mission in its Q2 data
release has a relatively small (1086) number of M dwarfs but their
data quality is much higher than for ground based programmes.  The
mission also samples uniformly in time with no large data gaps. We
briefly discuss the planet occurrence rate in this sample in Section
\ref{sec:comparison} based on \citetalias{Howard_2012} and compare it
with our results.
\par Compared to these other transit surveys, the WTS is targeting a
reasonable sample, even using only one of the four fields (at the time
of writing three of the four fields are lagging in coverage). The PTF
and Pan-Planets samples are potentially ground-breaking, if they can
obtain enough observations.
\subsection{A lack of hot Jupiters discovered around M dwarfs in the
WTS}
\label{sec:lackofplanets}
\par In the WTS, the initial candidate selection criterion is
currently a well-defined decision based solely on the signal-to-noise
ratio. Exhaustive eyeballing and photometric/spectroscopic follow-up
of the small number of candidate planets around M dwarfs
\citep{Sipocz_2013} has revealed them all to be grazing eclipsing
binary systems, or other false-positives. This search was complete for
all $\sim$Jupiter-sized candidates down to J=17 in the 19 hour field,
although avoiding objects with periods very close to one day. Two hot
Jupiters have however been discovered around more massive hosts
\citep{Cappetta_2012,Birkby_2013}. This confirms that the WTS
is sensitive to transits at the $\sim1$ per cent level (and indeed
these two objects are very obvious). \cite{Zendejas_2013}
have independently produced lightcurves (using difference imaging) and
searched for candidate planets in the same data, using stricter
automated classification, and also find no candidate Jupiters around M
dwarfs in the 19 hour field. This lack of hot Jupiters discovered
around small stars in the WTS is what motivates the rest of this
study, to address the question: how significant is this result.
\section{Simulation method}
\label{sec:simulation}
\subsection{Recovery ratio}
\par We determine the sensitivity of the WTS for four distinct
star-planet scenarios. For the two M dwarf subsamples, we consider
Neptune and Jupiter size planets in short period (0.8--10 days) orbits
around early (M0-2) and later type (M2-4) stars.  The mass and radius of
the star and the radius of the planet are kept fixed in each scenario
(see Table \ref{tab:systems}). We use conservative assumptions for the
stellar parameters: stellar radii are overestimated, using the maximum
values in each spectral bin (i.e at M0 and M2), and planet radii are
(under)estimated assuming solar system radii even for hot planets.
\begin{table}
\begin{center}
\begin{tabular}{>{\small}c@{\hskip .5em}>{\small}c@{\hskip .5em}>{\small}c@{\hskip .5em}>{\small}c@{\hskip .5em}>{\small}c@{\hskip .5em}>{\small}c@{\hskip .5em}>{\small}c}
\hline
No. & Sp.bin & $T_\mathrm{eff}$ (K) & $M_{*}/M_\mathrm{\odot}$ &
$R_{*}/R_\mathrm{\odot}$ &
$M_\mathrm{p}/M_\mathrm{Jup}$ &  
$R_{p}/R_\mathrm{Jup}$ \\
\hline
1 & M2-4 & 2960--3400 & 0.4 & 0.4 & 1.0 & 1.0 \\
2 & M2-4 & 2960--3400 & 0.4 & 0.4 & 0.054 & 0.36 \\
3 & M0-2 & 3400--3800 & 0.6 & 0.6 & 1.0 & 1.0 \\
4 & M0-2 & 3400--3800 & 0.6 & 0.6 & 0.054 & 0.36 \\
\hline
\end{tabular}
\end{center}
\caption{Simulated planetary systems\label{tab:systems}}
\end{table}
The expected number of recovered planetary systems can be written as:
\begin{equation}
N_\mathrm{det}= N_\mathrm{stars} f P_\mathrm{det}
\label{eq:expdet}
\end{equation}
where $N_\mathrm{stars}$ denotes the number of stars in the actual
star subtype group, $f$ is the (unknown) fraction of the stars that
harbour a planetary system, $P_\mathrm{det}$ is the (average)
probability of discovering a system by the survey around one of its
targets if we assume that the star harbours a (not necessarily
transiting) planetary system. $P_\mathrm{det}$ is expressed as a
function of planetary radius ($R_p$) and orbital period ($T$). In
general, we can write \citep{Hartman_2009}:
\begin{equation}
  P_\mathrm{det}=\int\int P_r P_T p(R_p,T) dR_p dT 
\label{eq:pdet}
\end{equation}
where $P_r$ is the average recovery ratio, i.e. the average
(conditional) probability of recovering a transit from a lightcurve if
the lightcurve belongs to a transiting system, $P_T$ is the geometric
probability of having a transit in a randomly oriented planetary
system, and $p(R_p,T)$ is the joint probability density function of
$R_p$ and $T$ for planetary systems. We determine these terms in Eq.\
\ref{eq:pdet} separately in each scenario. See also \cite{Burke_2006,
  Hartman_2009}.
\par We consider circular orbits only ($e$=0). $P_T$ can
be given analytically, for a random system orientation, $P_T = ( R_{*}
+ R_p )/a$ where $a$ denotes the semi-major axis of the system,
$R_{*}$ is the stellar radius. The joint probability density,
$p(R_p,T)$, must be treated as a prior. We consider planetary
configurations at discrete $R_\mathrm{P}$ values only, so the
$R_\mathrm{P}$ dependence of the density function simplifies to a
$\delta$-function. As the period dependence is ill-constrained, we
will discuss uniform and power-law functions as a prior
distributions. In the \citetalias{Howard_2012} study, a power law
model with an exponential cutoff at short periods was fitted (Table 5
in \citetalias{Howard_2012}) to Kepler giant planet detections around
mostly GK dwarfs (see sample criteria in Section
\ref{sec:comparison}). We use their model function normalized to our
studied period range as prior distribution.
\par $P_r$ is determined numerically by the Monte Carlo iteration
loop.
\subsection{Determining $P_r$}
\par We identify a set of 4700 {\em quiet} lightcurves in the 19hr
field that serve as input for the simulations. This {\em simulation
  sample} consists of mostly M dwarfs (and slightly hotter stars)
covering the magnitude range 11--17. The sample preserves the observed
distribution of M dwarf apparent magnitudes in the WTS. By quiet we
mean that the unperturbed lightcurves show no signature of a transit
(BLS $S_{\rm red}<=6$). By adding noise free signals to these
lightcurves and recovering them from the noisy data, we can quantify
the effect of noise on $P_r$. As shown in Section
\ref{sec:mdwarf_rms}, we found little difference between the noise
properties of the lightcurves for the M0-2 and M2-4 subclasses.
We simulate large numbers of transiting exoplanet systems to determine
the recovery ratios for the four scenarios under investigation. In
each iteration, a transiting planetary system is created with
parameters randomly drawn from fixed prior distributions as detailed
below (assuming a circular orbit). A simulated transit signal is then
added to the randomly selected (real) lightcurve. We try to recover
the artificial system using the transit detection algorithm discussed
in Section\ \ref{sec:transdet} (\cite{Aigrain_Irwin_2004}). $P_r$ is
estimated as the ratio between {\em successful} transit recoveries and
the total number of iterations. We make a distinction between two
different cases. In the {\em threshold} case we consider the signal
successfully recovered if the detection passes the same signal to red
noise level as used for WTS candidate selection
($S_\mathrm{red}$>6). In the {\em periodmatch} case, we additionally
require that the recovered period value matches the simulated
one. This is discussed in more detail in Sec. \ref{sec:det_period}.
\par The period value ($T$) is drawn from a uniform distribution in
the range of $0.8$ to $10$ days.  The period determines the semi-major
axis ($a$) of the system, given the masses of the star (assumed to be
0.6 or 0.4$M_\odot$) and the planet (assumed to be 1.0 or 0.054
$M_\mathrm{Jup}$). A randomly oriented system is uniformly distributed
in $\cos i$ where $i$ is its orbital inclination. The random
inclination is chosen to satisfy $0\leq\cos i < (R_{*} + R_{p})/a$ to
yield a transiting system. This also allows for grazing
orientations. The phase of the transit is also randomly chosen from
within the orbital period.
\par Observed dates in the target lightcurve are now compared with
predicted transit events. If there would be no affected observational
epochs, then the iteration ends and the generated parameters
recorded. Otherwise, a realistic, quadratic limb darkening model is
used with coefficients from \cite{Claret_2000} to calculate brightness
decrease at in-transit observational epochs
\citep{Mandel_Agol_2002,Pal_2008}. This artificial signal is added to
the lightcurve magnitude values, and BLS is run on the modified
lightcurve. Both generated and detected transit parameters are
recorded for the iteration. We note that the transit detection
algorithm is the most computationally intensive step in the loop. A
total of 75,000 iterations were performed.
\section{Results and Discussion}
\label{sec:discussion}
\subsection{Sensitivity effect of the observation strategy}
\par
\begin{figure}
  \centerline{\includegraphics[width=9cm]{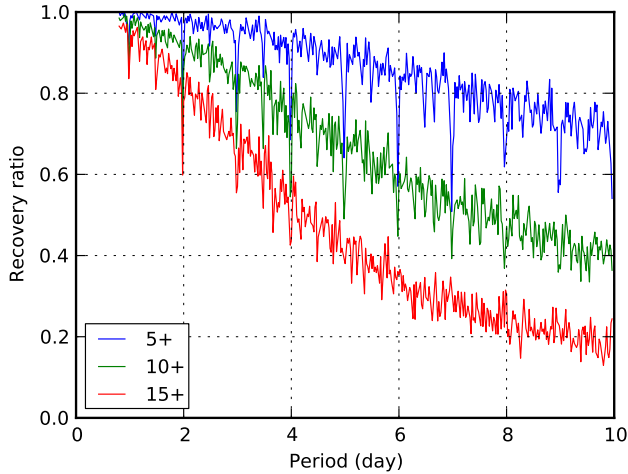}}
  \caption{The effect of the observation strategy on the WTS
    sensitivity to short period transiting systems in the 19 hour
    field. We use the actual epochs of our 19 hour field observations
    combined with a large sample of simulated planets. The fraction of
    simulated transiting systems is shown where at least 5, 10 or 15
    {\em actual} WTS observational epochs coincide with the {\em
      simulated} in-transit times. This illustration does not use the
    noise-properties of the WTS, thus does not consider the case as to
    whether the events are actually detected or not.}
  \label{fig:basic_sens}
\end{figure}

 The flexible observing mode of WTS has an inherent limitation on
our sensitivity to short-period transiting systems, and it takes
multiple seasons to build up enough epochs to reliably detect them. In
Fig.\ \ref{fig:basic_sens} a simple sensitivity diagram is shown which
considers only the actual distribution of observational epochs for the
19 hour field. We use the simulated transiting systems from the
Neptune-size planets around an M0 dwarf scenario and compare the {\em
  simulated} transit times with our real observational
epochs. A system is considered detectable here if at least 5,
10 or 15 in-transit observational epochs occur calculated from the
{\em simulated} parameters. The fraction of detectable systems has an
obvious strong dependence on the period of the transiting system and
the required number of in-transit observational epochs as well.  Of
course, it depends on the noise properties of our data how many
in-transit observations are necessary for detecting a transit
event.
\subsection{Detection statistic}
\label{sec:det_stat}
\par
\begin{figure*}
  \centerline{\includegraphics[width=9cm]{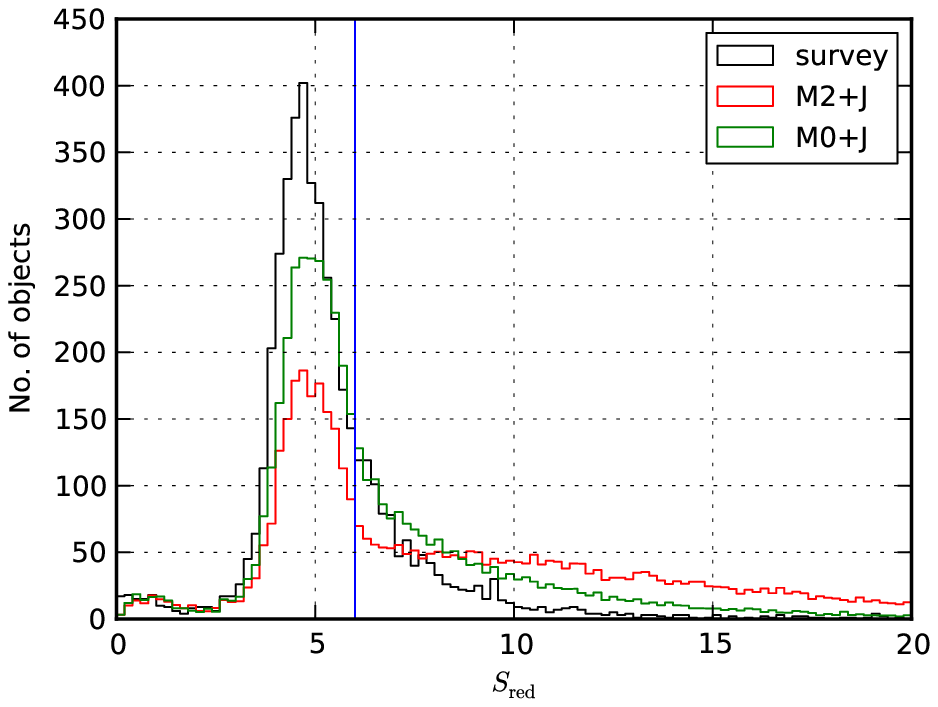}
  \includegraphics[width=9cm]{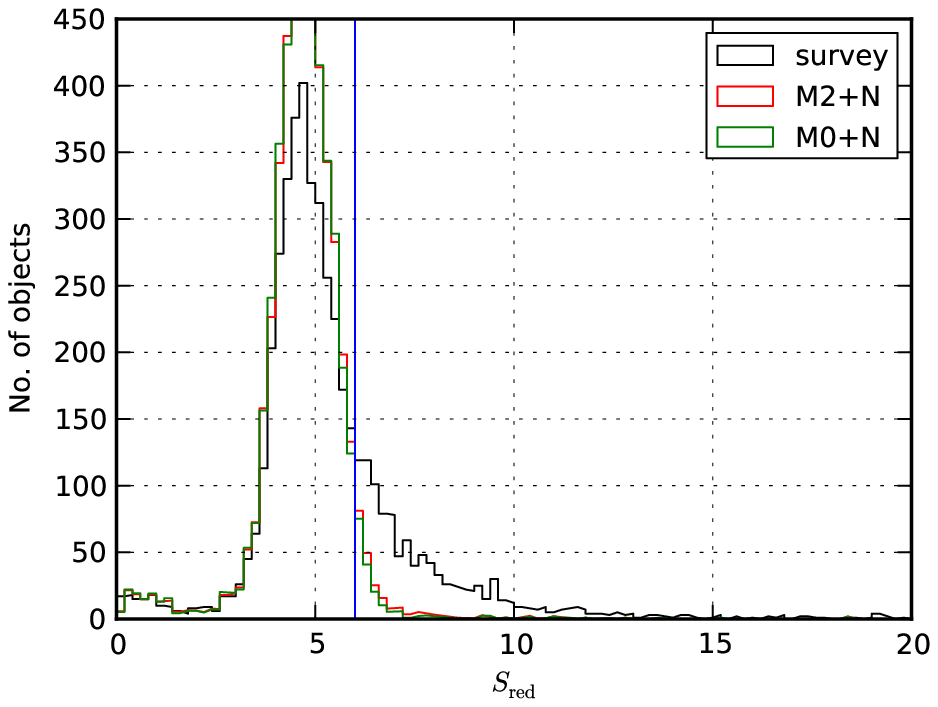}}
\caption{Detected signal to red noise histograms (bin
    size is 0.2) of the transit injected quiet lightcurve iterations
    (coloured) for Jupiter (left) and Neptune (right) size planets in
    the four simulated scenarios. For comparison, the unmodified
    (black) M dwarf lightcurves of the survey is shown and curves of
    the iteration loops are normalized to the number of M dwarf
    objects in the survey (black curve). The vertical line indicates
  our candidate selection threshold of 6.}
  \label{fig:sred_histogram}
\end{figure*}

 We show the (signal to red-noise) detection statistic
distributions in Fig.\ \ref{fig:sred_histogram}. The black curve
belongs to the survey M dwarfs (unmodified data) in the 19hr field
(selection criteria described in \ref{sec:mdwarf_bins}) . The coloured
curves are for the simulated transiting systems for the four
scenarios. For comparison, they are normalized to the total number of
M dwarfs in the survey, i.e as if every WTS M dwarf target were either
an M0 or an M2 with a transiting Jupiter or a Neptune (which may, or
may not, have transited during the WTS observations).
\par The original (unperturbed) WTS M dwarf $S_\mathrm{red}$
distributions shows a marked tail above the detection threshold (25
per cent of all objects, see Sec.\ref{sec:false_pos}). This could be
caused by effects such as: correlated noise, intrinsic variability
(spots), eclipsing binaries or possibly transiting planets. The
simulations show that for initially ``quiet'' lightcurves
($S_\mathrm{red}<6$) the injection of a planet signal (transiting, but
could be grazing) can perturb the measured signal-to-noise value above
the detection threshold. (M2+J:50, M0+J:39, M2+N: 5.1, M0+N: 4.1 per
cent of all iterations) In other words, the WTS is sensitive to
Jupiters, and rather less sensitive to Neptunes. Given the other
possible causes for a perturbed $S_\mathrm{red}$, the shape of this
distribution does not in itself present a direct measurement of the
planet population. It is worth noting that for the Jupiters
(M0+J,M2+J), the recovered $S_\mathrm{red}$ values can be rather
higher than the largest values we actually see in the WTS.
\par The Neptune cases have much smaller residual tails which sit only
a little above the detection threshold. These signal injections cause
detection statistic increases above the detection threshold only in a
small number of cases. There is also little dependence on the size of
the host star (the green and red curves are similar). Our simulations,
with the same stellar magnitude distribution as the survey, are
dominated by fainter objects. The similarity of the detection
statistic distributions in the figure indicates that for Neptune sized
planets, the survey sensitivity depends on the stellar radius only for
the brightest stars. For fainter objects, the sensitivities are the
same and as we see later, they are equally low (Figures
\ref{fig:rec_ratio_m0} and \ref{fig:rec_ratio_m2}).
\subsection{Detected period}
\label{sec:det_period}
\par While passing the signal-to-noise statistic threshold in the
simulation is caused by the injected signal, the detected (recovered)
parameters, particularly the period, may not be correct. Indeed, it is
common in ground based surveys to find multiple peaks in the BLS
periodogram containing aliases of the observing window function.
\par The detected period is perhaps the most important of the model
parameters: At the eyeballing stage of candidates, this period has the
most influence in judging a transit detection real or false. In a
lightcurve folded on the correct period (or a harmonic thereof), the
transit signals are visually easily recognized. When folded on a
random period, in-transit points are hardly distinguishable from 
random outliers. A good initial period value is also important for the
timing of follow-up observations. With this in mind, we try to account
for the importance of this effect in our simulations. We consider two
recovery rates:
\begin{itemize}
\item {\em threshold} --- $S_\mathrm{red}$ exceeds 6.0, and 
\item {\em periodmatch} --- we additionally require that the detected
period value matches (or be a harmonic of) the generated period. We
allow factors of 1, 5/4, 4/3, 3/2, 5/3, 2, 5/2, 3, 4, 5 between the
two values. 
\end{itemize}
\par This step uses {\em external} information and obviously we cannot
know the true period in the actual survey. However, in the
simulations, it does allow us to place more strict criteria on
detected simulated planets, and mimic the effects of differentiating
between high priority candidates, which are likely to receive
follow-up time, and low-priority candidates for which follow-up may be
too expensive. This effect is much more significant for the Neptunes,
where individual events are shallow.
\par
\begin{figure*}
\centerline{\includegraphics[width=9cm]{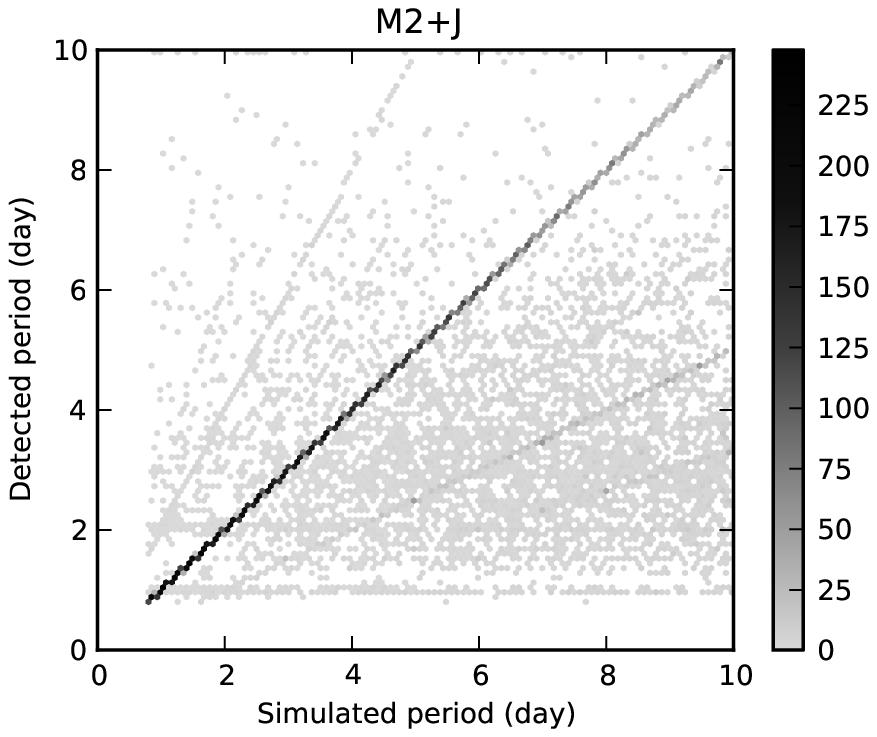}
  \includegraphics[width=9cm]{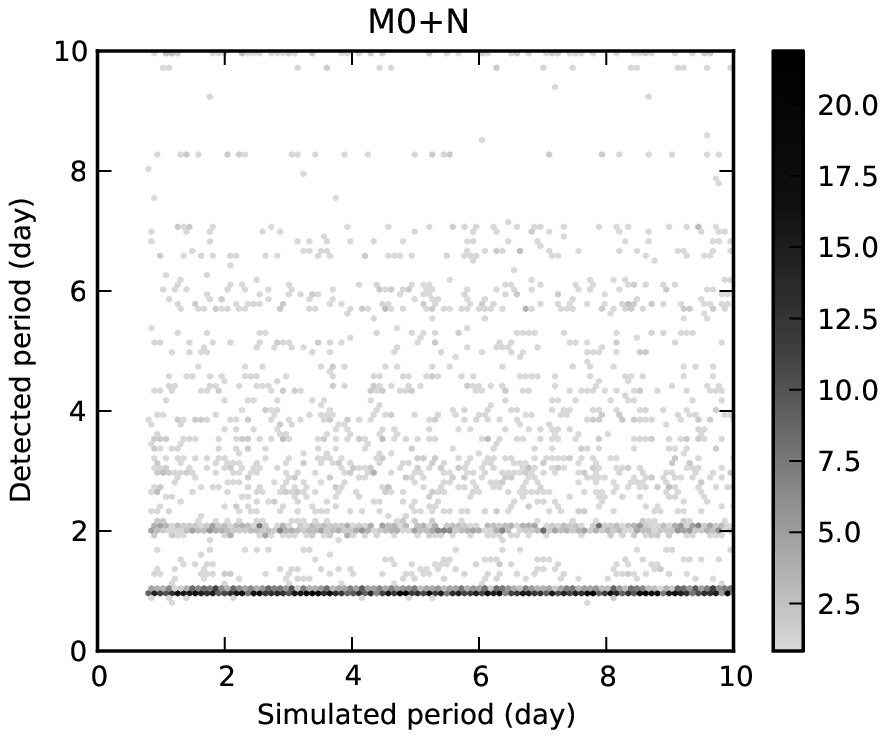}
}
\centerline{\includegraphics[width=9cm]{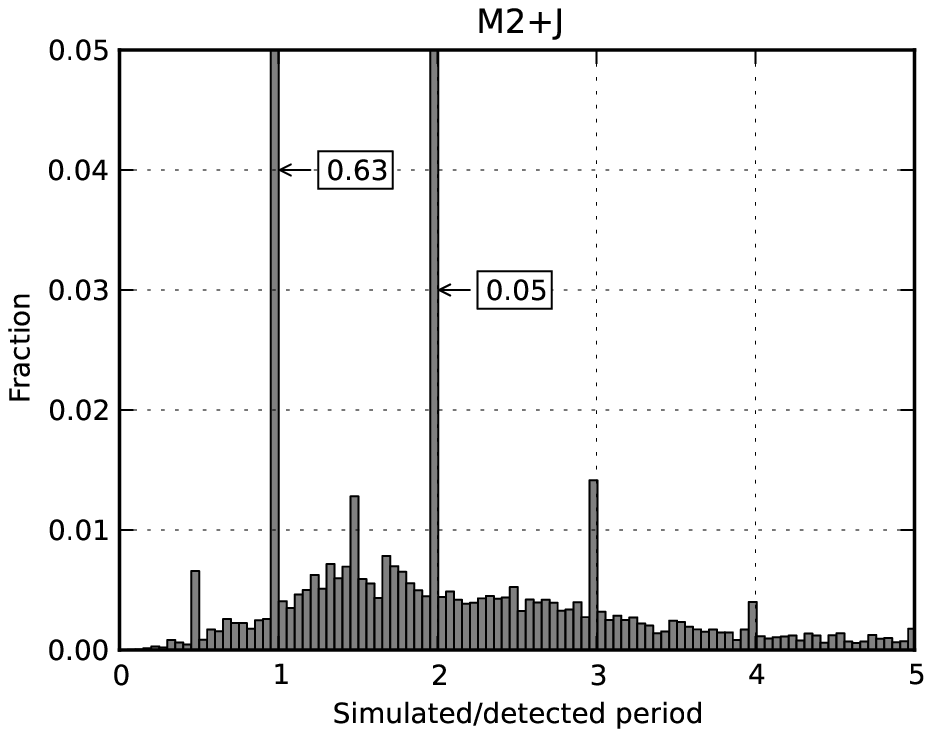}
  \includegraphics[width=9cm]{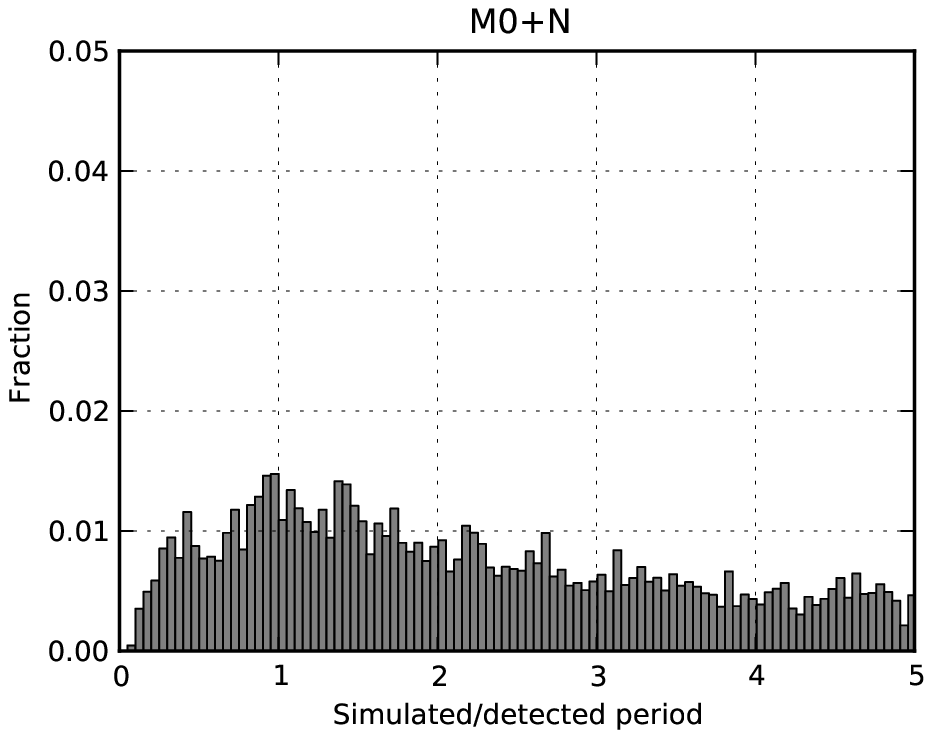}  
}
\centerline{\includegraphics[width=9cm]{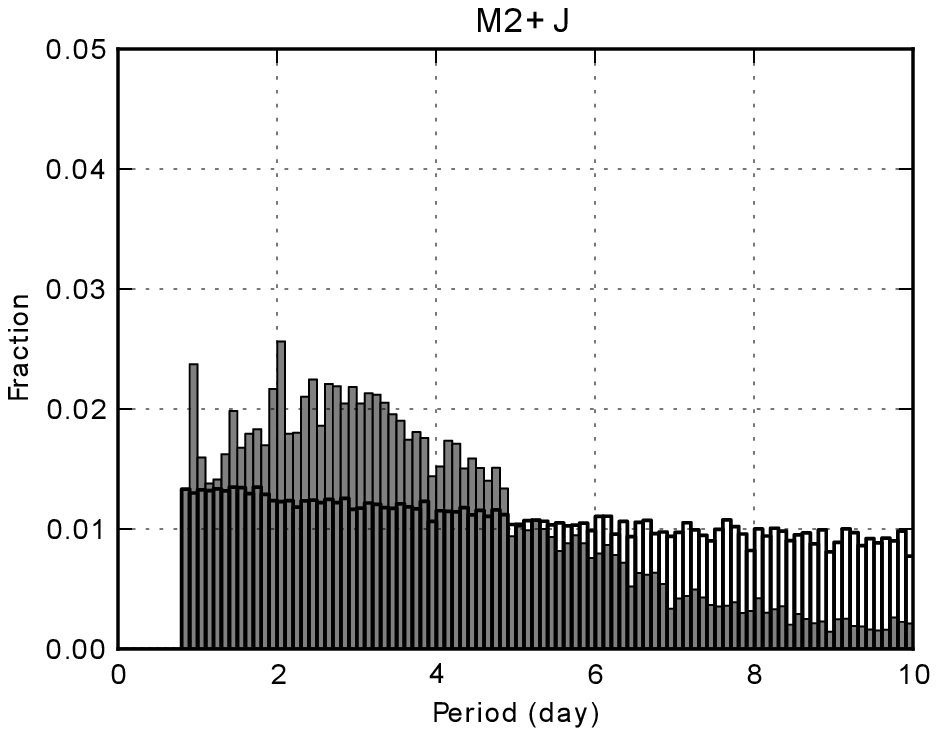}
  \includegraphics[width=9cm]{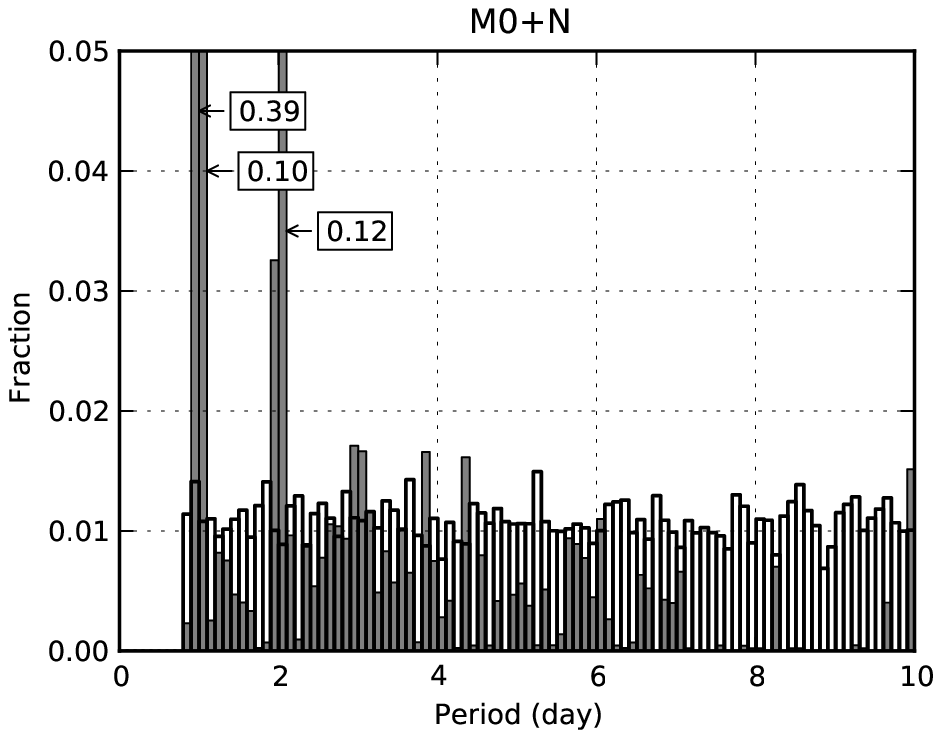}
}
\caption{Properties of period values of iterations that pass the
  signal-to-noise threshold in the most (M2+Jupiter, left) and least
  (M0+Neptune, right) sensitive scenarios. Best fit detected period
  values as a function of simulated period values (top row) and
  normalized histograms of simulated/detected period ratios (middle
  row; bin size is 0.05). Bottom row: Normalized histograms (bin size
  0.1) of detected (filled) and simulated periods (empty)
  i.e.\ projections to axes of top row scatter panels.}
\label{fig:detp_simp}  
\end{figure*}

 We note that the number of in-transit data points is correlated
with recovering the correct period. In the M2+J case, we find that if
a simulated system has 10 in-transit data points we detect the correct
period 80 per cent of the time, reaching almost 100 per cent with 20
in-transit data points. For larger stars (shallower transits), more
in-transit data points are needed (M0+J: 12 and 30 points for 80 and
100 per cent respectively). For Neptunes even more transits are
required to secure the period. For the M2+N case, we find 20
in-transit points recovers 50 per cent of the detected systems with
the correct period. For the M0+N case the situation is even worse, and
we almost always detect an alias (see Fig.\ \ref{fig:detp_simp}). This
confirms that the BLS algorithm is ``lost'' at these low signal
levels. Comparing the necessary number of in-transit data points for
reliable transit detections, according to Fig.\ \ref{fig:basic_sens}, we
can see an inherent limitation of the transit searching efficiency of
the WTS and possibly other ground based low cadence surveys
(e.g. PTF).
\par In this study we determine $P_r$ for the {\em threshold} and {\em
  periodmatch} cases. The real value of $P_r$ that characterizes the
WTS probably lies between these two extremes and depends on all the
(sometimes subjective) steps of the candidate selection and follow-up
strategy. Having fewer quality criteria in the survey increase the
number of candidates at the cost of higher false alarm ratio and more
follow-up resource consumption.
\par We note that surveys with close to real time data processing and
instant access to follow-up observation facilities could use a
different approach and need not necessarily balance between quality of
recovered transits and false positive rates. They can maximize the
probability of detecting actual transit events by selecting follow-up
times optimized to their current data and iteratively revise
predictions as data accumulates. A Bayesian approach to such a
strategy is described in \cite{Dzigan_2011}.
\par In Fig.\ \ref{fig:detp_simp} scatter diagrams and histograms of
simulated and detected periods are shown for recovered simulated
planets. We compare our most sensitive case in terms of injected
signal depth (M2+J), to the least sensitive case (M0+N). They show a
clear contrast in the quality of detected (signal to noise selected)
transit signals.
\par For Jupiters, about 70 per cent of the detections also recover
the injected period or its harmonic value (middle row in Fig.
\ref{fig:detp_simp}).  Strong harmonic lines can be seen in the
scatter panel (top left). For M0 stars (not shown), the ratio of
recovered period values is even better. The shallower transits are
longer in duration for a given period. The detected periods (bottom
row, filled histogram) are biased towards shorter values peaking
around 3 days, with aliases appearing at 1 and 2 days. In these
iterations the injected signal changes the lightcurve in a way that
aliased periods give the most significant box-fit signals. We see a
gently decreasing trend for the periods of simulated systems that were
recovered (bottom row, empty histogram), i.e. we are less sensitive to
longer period planets, and more likely to underestimate their
periods. We can conclude that signals of Jupiter sized planets can
firmly be detected around all of the M dwarfs in the survey with good
initial period values.
\par The Neptunes are shown in the right-hand column of
Fig.\ \ref{fig:detp_simp}. In accordance with our qualitative
assessment based on the lightcurve RMS, signals of Neptune sized
planets are at the boundary of being lost in the lightcurve noise in
the WTS. In the M2+N case, only 12 per cent of the detections have
also matching periods, in the M0+N case the correct period is almost never
recovered by the box fitting algorithm (middle row). Recovered period
values are heavily dominated by alias values. We conclude that Neptune
sized planets can be detected in the survey only in favorable cases
around smaller (later) M dwarfs.
\subsection{False positives}
\label{sec:false_pos}
In a transiting planet survey, we have to deal with two types of false
positives at the initial transit candidate discovery step. The
detection statistic may pass where the transit box model is fitted
just on (i) random fluctuations or systematics. In this case, the
transit signal does not exist at all. (ii) Physical signals in the
lightcurve may also belong to various eclipsing binary configurations
or to variable stars (e.g.\ from spots) which can mimic transit events
for the algorithm.
\par The amount of additional analysis to rule out false positive
candidates can vary from little additional manual checking, through
refinement of transit parameters, up-to obtaining additional
observational data with better cadence. Grazing binary stellar systems
can mimic planetary transits beyond the survey's photometric precision
and require additional spectroscopy measurements to be ruled out with
high confidence.
\par The false positive ratio of the candidate selection procedure of
the survey cannot be quantified by the simulation alone. A lightcurve
passing the detection threshold with known injected simulated transit
signal cannot be a false positive per se.
\par So far, all the actually selected, eyeballed and later
followed-up (by additional photometry and/or spectroscopy) candidates
around M dwarfs turned out not to be a planetary system
\citep{Sipocz_2013}. Following \cite{Miller_2008}, this can be used
to estimate the false positive ratio of the survey. This encompasses
both false positive cases above.  We assume that all
$S_\mathrm{red}>6$ detections in our original (unmodified) lightcurves
are false positives. At 25 per cent (black curve above the marker in
Fig.\ \ref{fig:sred_histogram}), this is a rather large fraction. This
number gives a simple (worst-case) description of our false positive
ratio, and cannot be usefully applied to the simulations.
\subsection{Transit recovery ratios}
\label{sec:recovery_ratio}
\par
\begin{figure}
\centerline{\includegraphics[width=9cm]{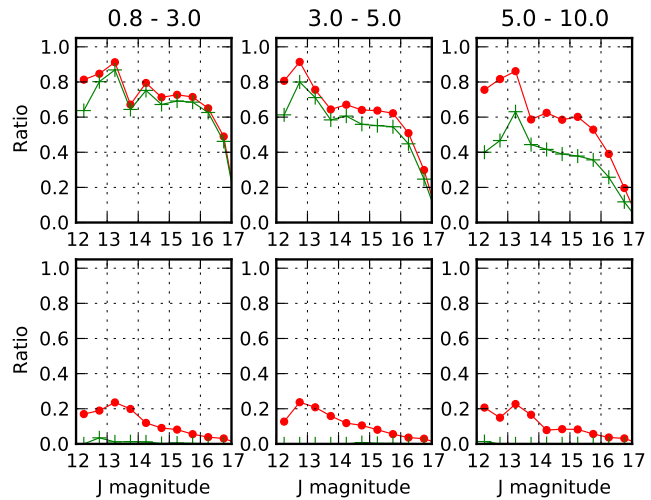}}
\caption{Recovery ratios ($P_r$) of simulated transiting system for
  M0+Jupiter (upper row) and M0+Neptune (lower row) scenarios in three
  period ranges as a function of stellar brightness. Filled circles
  (\textbullet) and crosses (+) represent the {\em threshold} and
  {\em periodmatch} ratio respectively.}
\label{fig:rec_ratio_m0}
\end{figure}

\begin{figure}
\centerline{\includegraphics[width=9cm]{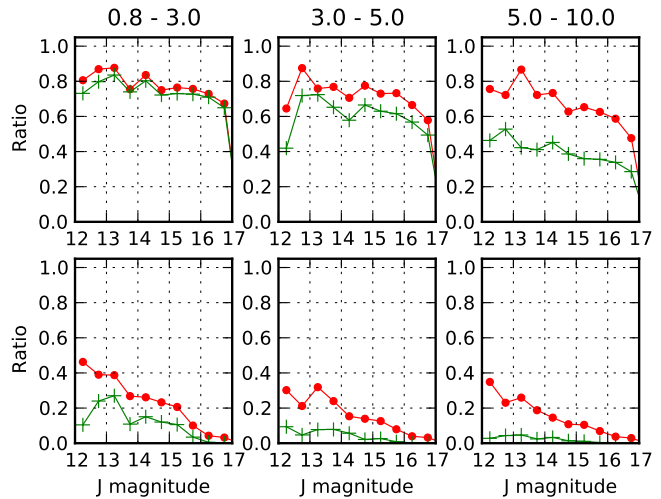}}
\caption{Recovery ratios ($P_r$) for M2+Jupiter (upper row) and
  M2+Neptune (lower row) simulated scenarios.}
\label{fig:rec_ratio_m2}
\end{figure}

 We show results of $P_r$ as a function of stellar magnitude for
ranges of simulated period in Figs.\ \ref{fig:rec_ratio_m0} and
\ref{fig:rec_ratio_m2}. We adopt 0.8--3.0, 3.0--5.0, 5.0--10.0 day
ranges following \cite{Hartman_2009} for extremely hot Jupiters (EHJ),
very hot Jupiters (VHJ), and hot Jupiters (HJ) respectively. Filled
red circles (\textbullet) and green crosses (+) represent the {\em
  threshold} and {\em periodmatch} cases respectively. Integrated
detection probabilities ($P_\mathrm{det}$) weighted by different prior
assumptions including the geometric probabilities of transiting
orientations are shown in Table \ref{tab:frac_planets}.
\par For the Jupiter cases (upper rows in both figures), the WTS
sensitivity has a maximum around J=13.5 and drops towards fainter
(J>15-16) objects. This is in accordance with our expectations, we
have higher noise levels towards fainter objects but there is
occasional saturation at the bright end ($J<13$). There is little
dependence on stellar radius, appearing only for the fainter
stars. The {\em threshold} curves are less affected by simulated
period than the {\em periodmatch} recovery rate. In the shortest
period window the {\em threshold} and {\em periodmatch} values are
practically the same. In the longest period panels the {\em threshold}
ratio is about twice that of the {\em periodmatch} ratio showing that
while the signal-to-noise detection statistic can recover signals,
many of these systems may be missed due to poor initial period
guesses.
The survey has a much lower sensitivity for Neptunes (bottom rows in
Figs.\ \ref{fig:rec_ratio_m0} and \ref{fig:rec_ratio_m2}). The {\em
  threshold} curves show roughly 25 per cent of the recovery rate seen
for Jupiters around bright stars. The difference between the {\em
  threshold} and {\em periodmatch} case is also more significant. The
{\em periodmatch} recovery rates are close to zero for all M0 cases
and very low for the M2 cases as well. The only exception is the best
signal-to-noise case around M2 stars, in the short period window for
bright objects where {\em threshold} rates reach half of the Jupiter
value and the {\em periodmatch} rates are not close to zero. It looks
like the WTS survey only really has a chance of discovering extremely
hot Neptunes around late (M2 or later) and bright M stars
($J<15$). The lack of an accurate period determination would make
follow-up of Neptune size candidates difficult in all other cases.
\subsection{Error considerations} 
\par We use the seeing corrected lightcurves as a source for our
simulation. While it could be more accurate to modify measured raw
flux values and to run all the pipeline components to create the
modified lightcurves, it would be unreasonably resource consuming, nor
assumed to have an impact on our sensitivity
results. \cite{Burke_2006} decrease the sensitivity with a constant
estimation of $0.03$ to deal with their trend filterings' flattening
effect. In our case it is assumed to have a negligible difference only
as we do not use trend filters that uses solely lightcurve data. As it
is described in the pipeline overview, images are transformed into
lightcurves simultaneously, calculating magnitude scale offsets using
non-variable, bright objects on each frame. Compared to the several
thousand objects per frame, only a small fraction of objects may have
a transit on a given frame thus the determined magnitude offset value
would be independent of an injected signal. Being a bad weather
fallback project, the survey typically has a few (2-3) observational
epochs per observing night. Thus we do not expect significant correlation
between the seeing and the timing of transit signals.
\begin{table}
\begin{tabular}{ccccccc}
\hline
System type & prior & $N_\mathrm{stars}$ & $P_\mathrm{det}$ & $f_\mathrm{95\%}$ & 
$P_\mathrm{det}$ & $f_\mathrm{95\%}$ \\
&&&sn&sn&pm&pm\\
\hline
\hline
M0+Jupiter &Kep. & 2844 & 0.0363  & 2.9\% & 0.0311 & 3.4\% \\
M2+Jupiter &Kep. & 1679 & 0.0414  & 4.3\% & 0.0346 & 5.2\% \\
M0-4+Jupiter & Kep. & 4523 & 0.0382  & 1.7\% & 0.0324 & 2.0\% \\
%
M0+Jupiter &uni. & 2844 & 0.0360 & 2.9\% & 0.0305 & 3.5\% \\
M2+Jupiter &uni. & 1679 & 0.0405  & 4.4\% & 0.0332 & 5.4\% \\
M0-4+Jupiter &uni. & 4523 & 0.0377 & 1.8\% & 0.0315 & 2.1\% \\
\hline
M0+Neptune &Kep. & 2844 & 0.0027 & 39\% & 0.0001 & 100\% \\
M2+Neptune &Kep. & 1679 & 0.0025 & 71\% &  0.0003 & 100\% \\
\hline
\end{tabular}
\caption{Planet detection probability and upper limit results on
  planetary fractions in the WTS in the signal-to-noise {\em
    threshold} (sn) and {\em periodmatch} (pm) interpretation. The
  upper part shows limits for the simulated Jupiter scenarios (M0,M2)
  and its interpretation for the whole M0-4 dwarf sample. Values are shown
  for priors based on the \citetalias{Howard_2012} study (Kep.) and
  uniform (uni.). The lower part shows representative values for the
  Neptune cases.\label{tab:frac_planets}}
\end{table}
\subsection{Limits on planetary fraction}
Assuming that the survey planetary candidates are all false positives,
as our follow-up suggests (see Sec.\ \ref{sec:lackofplanets}), we can place
upper limits on the planetary fraction ($f$) for our four test
cases. The assumption of complete follow-up is much less secure for
Neptune-sized planets, nevertheless the upper limits remain valid for
the survey until a signal is detected. While Eq.\ \ref{eq:expdet}
gives the expected number of detections, the actual number of
detections follows a Poisson distribution. If the expected number of
planets is $N_\mathrm{det}$, the probability of detecting $k$ planets
is:
\begin{equation}
P_k=\frac{N_\mathrm{det}^k}{k!}\exp(-N_\mathrm{det})
\end{equation} 
We use this expression at $k=0$ as a likelihood function for
$N_\mathrm{det}$. We require that $N_\mathrm{det}$ be within our $0\le
N_\mathrm{det}<N_\mathrm{max}$ confidence interval with 95 per cent
confidence\footnote{This is actually a Bayesian interpretation for the
  95 per cent probability credible interval for the parameter of the
  Poisson distribution, assuming a sufficiently wide, uniform prior
  for the parameter. For nonzero cases in Sec.\ref{sec:comparison},
  intervals with equal posterior probability values at the endpoints
  are chosen.}. Solving
\begin{equation}
P(N_\mathrm{det}<N_\mathrm{max}) = \int\limits_0^{N_\mathrm{max}}
\exp(-N) dN = 0.95
\end{equation}
for $N_\mathrm{max}$, we get $N_\mathrm{max}= 3.0$. From the
requirement of $N_\mathrm{det}<3.0$, using Eq.\ \ref{eq:expdet}, we
get:
\begin{equation}
f_{95\%} \le \frac{3.0}{N_\mathrm{stars} P_\mathrm{det}}
\label{eq:uplim}
\end{equation}
\par Assuming zero detections and applying the above technique, we can
compute robust upper limits on the fraction of planet host stars in
our sample. We show the {\em threshold} and {\em periodmatch}
$P_\mathrm{det}$ values along with the planetary fractions in Table
\ref{tab:frac_planets}. We show results for the two different
assumptions for the period distribution: best fit exponential cut
power law from Kepler \citepalias{Howard_2012} (Kep.), uniform (uni.).
For the hot Jupiters as a whole (periods 0.8--10 days), we can place
an upper limit of $f_{95\%}=$2.9--3.4 per cent planet fraction around the early M
subtype group (M0-2) assuming the Kepler power law prior period
distribution. Using the simple uniform prior gives similar
sensitivities (2.9--3.5 per cent) in this short period range. The
smaller sample of cooler stars leads to a higher upper limit of
4.3--5.2 (4.4--5.4 for uniform) per cent for the M2-4 group.  For
Jupiter size planets, the {\em threshold} and {\em periodmatch}
detection probabilities are essentially the same, and the chosen prior
period distribution has also little impact.
\par We can combine the M0--M4 stars into one bin by calculating the
average sensitivity for the whole sample weighted by the number of M
dwarfs in the corresponding groups. The WTS can put an upper limit of
$f_{95\%}=$1.7--2.0 (1.8--2.1 for uniform) per cent on the occurrence rate of
short period Jupiters around early-mid (M0-4) M dwarfs. We discuss
these results in context in Section\ \ref{sec:comparison}.
\par For Neptunes, the WTS detection probabilities are an
order-of-magnitude smaller than for Jupiters. We also find a strong
dependence both on the assumed prior for the period distribution and
the {\em threshold}/{\em periodmatch} interpretations. As seen,
the recovery of Neptunes in the WTS is a difficult and unreliable task.
The probability of recovering a Neptune with the correct period ({\em
  periodmatch} $P_\mathrm{det}$) is so low because detected signals
are dominated by aliased periods. Thus it is very difficult to
distinguish genuine Neptunes from false detections. The main way to
improve on the WTS sensitivity to Neptunes would be to improve on the
noise properties in the data. This is beyond the scope of this paper.
\section{Discussion: Comparison with Kepler and RV studies}
\label{sec:comparison}
Although Kepler is not a specialized survey for M dwarfs, due to its
exceptional quality data and large number of targets, it is probably
the highest impact exoplanetary transit survey to date. In this section we
compare our results to the planet occurrence study of the Kepler data
in paper \citetalias{Howard_2012}.
\par In the second part of the \citetalias{Howard_2012} study, the
planet occurrence rate is determined in three planet radius bins
(2-4$R_\oplus$, 4-8$R_\oplus$, 8-32$R_\oplus$) as a function of
stellar type (effective temperature; in 500K wide temperature bins
from 3600K to 7100K; see figure 8 in \citetalias{Howard_2012}). They
use the Q2 Kepler data release \citep{Borucki_2011}.  In each stellar
temperature bin, the total planet occurrence rate is calculated by
adding the contribution of each discovered (high quality) Kepler
planet (candidate). eq.2 from \citetalias{Howard_2012} is reproduced
here:
\begin{equation}
f=\sum_{j=1}^{N_\mathrm{pl}}\frac{1/P_T}{n_{*,j}}
\end{equation}
$f$ is evaluated for each studied stellar bin. Each planet's
contribution is augmented by its geometric transit probability
($1/P_T$) to include planetary systems with non-transiting
orientations as well (i.e.\ a detected planet with low geometric
transit probability give a high contribution). Each planet's
fractional contribution is calculated over the number of high
photometric quality stars only ($n_{*,j}$). Only those stars are
selected from the Kepler Input Catalog (KIC) that belong to the
analyzed stellar bin and have a high enough quality lightcurve where
the actual planetary transit can be certainly detected (Table
\ref{tab:kepler_prop}). A SNR value is derived from the edge-on
planetary transit depth signal and from the measured scatter
($\sigma_\mathrm{CDPP}$; see eq.1 in \citetalias{Howard_2012}) of the
lightcurve. A lightcurve counts in the total number of stars if SNR>10
is fulfilled. 
\par For transiting planets, a SNR>10 transit detection (calculated
from the actual transit signal depth) and orbital period $T<50$ days
are required. These planets are not all confirmed yet but part of the
released Kepler planetary candidates passing an automated vetting
procedure of the Kepler data processing pipeline. They are counted as
planets both in \citetalias{Howard_2012} and in the present paper as
they are assumed to be actual planets with high probability
\citep{Borucki_2011}.
\begin{table}
\begin{tabular}{rl}
\hline
Stellar effective temperature bins&3600--4100, ..., \\
& 6600--7100K\\
Stellar surface gravity, log g&4.0--4.9\\
Kepler magnitude, $K_p$&$<15$\\
Lightcurve quality, SNR& $>10$\\
\hline
Planetary radius bins, $R_\oplus$&2-4, 4-8, 8-32\\
Orbital period, $T$&$<50$ days\\
Detection threshold, SNR&$>10$\\
\hline
\end{tabular}
\caption{Properties of stellar and planetary
  samples considered in the \citetalias{Howard_2012} Kepler study.\label{tab:kepler_prop}}
\end{table}
\par To obtain stellar parameters, the Kepler project uses model
atmospheres from \cite{CK_2004}. They perform a Bayesian model fitting
on seven colours of the KIC objects determining effective temperature
($T_\mathrm{eff}$), surface gravity ($\log g$) and metallicity ($\log
Z$) among other non-independent model parameters simultaneously
\citep{Brown_2011}. Restrictive priors also ensure that parameters
remain within realistic value ranges. Temperatures are considered to
be most reliable for Sun-like stars with differences from other models
below 50K and up to 200K for stars further away from the Sun on the
CMD. \cite{Brown_2011} call temperatures below 3750K {\em
  untrustworthy}. There are 1086 M dwarfs identified in Q2 (Table
\ref{tab:kepler_jup}).
\par
\begin{figure}
\centerline{\includegraphics[width=9cm]{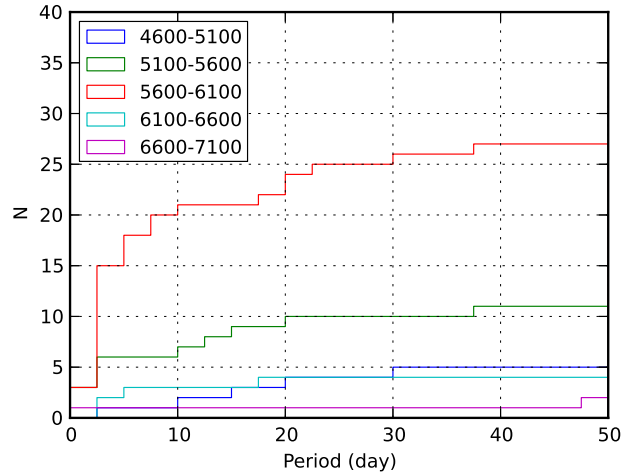}}
\caption{Cumulative number of discovered Jupiter size (8-32$R_\oplus$)
  planets in Kepler Q2 as a function of orbital period (T) in
  different stellar temperature bins. There are no planets in the two
  coolest stellar bins (3600--4100K, 4100K--4600K). Planet occurrence
  fractions in Fig.\ \ref{fig:kepler_frac} are calculated for the
  0.8--10 days region, augmenting the contribution of each discovery
  by its geometric transit probability.}
\label{fig:kepler_jupiter_NvsP}
\end{figure}

 Using the Q2 Kepler data release, we calculate Kepler planetary
fractions for the 0.8--10 days period range to compare with our
present WTS study. We follow the steps of \citetalias{Howard_2012} by
selecting KIC objects and Kepler planets (candidates) from the Q2 data
release using most criteria listed in Table \ref{tab:kepler_prop} but
selecting planets with $T<10\ \mathrm{days}$ (Fig.\
\ref{fig:kepler_jupiter_NvsP}). We cannot filter for high quality
lightcurves however as noise data is not readily available from
\citetalias{Howard_2012}. It is beyond the scope of this paper to
process each released Kepler lightcurve and check its noise
properties. Rather, we determine a correction factor for each stellar
type bin. We reproduce the calculation for the planetary fraction in
the $T<50$ day case omitting the lightcurve SNR>10
quality criterion and compare it to the published values in fig. 8 in
\citetalias{Howard_2012}. This reveals a correction factor that is
applied to the $T<10\ \mathrm{day}$ case in each stellar bin (Table
\ref{tab:kepler_jup}). 
\par
\begin{figure}
\begin{center}
\includegraphics[width=9cm]{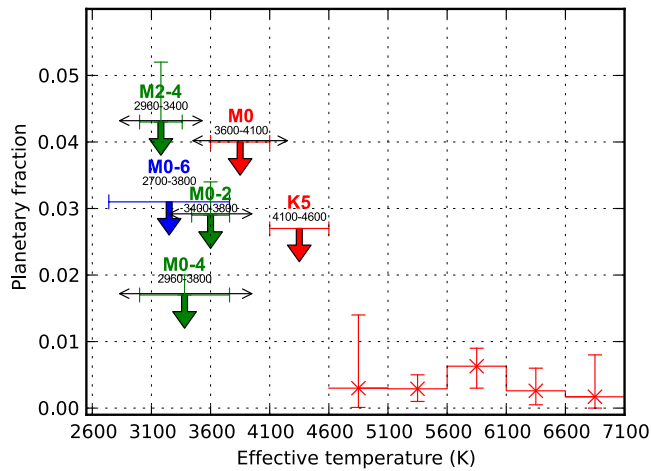}
\end{center}
\caption{Short period (0.8--10 day) hot Jupiter planetary occurrence
  fractions (x) and upper limits ($\downarrow$) in case of null
  detections in the WTS (green), Kepler Q2 \citepalias{Howard_2012}
  (red) and from the HARPS planet search \citep{Bonfils_2011}
  (blue). WTS upper limits are shown for the M0-2, M2-4 bins and also
  for the whole M0-4 sample (green arrows). Vertical error bars on WTS
  upper limit markers cover the uncertainty from the {\em threshold}
  and {\em periodmatch} interpretations (Table
  \ref{tab:frac_planets}). Horizontal bars ($\vdash\dashv$) show
  temperature bin widths, horizontal arrows ($\leftrightarrow$) mark
  estimated uncertainty of bin edges where available. All upper limits
  and error bars of nonzero fractions are for 95 per cent confidence.}
\label{fig:kepler_frac}
\end{figure}

 For stellar bins with nonzero planets 95 per cent confidence
intervals (red error bars in Fig.\ \ref{fig:kepler_frac}) are calculated
following the logic of `effective stars' from
\citetalias{Howard_2012}; using a Poisson distribution for having the
actual number of planet detections ($N_\mathrm{pl}$) from
$N_\mathrm{pl}/f$ `effective stars'. Most stellar bins have only a few
detections, so errors are heavily dominated by small number
statistics. For bins with zero candidates (95 per cent confidence)
statistical upper limits are calculated in the same way as for the WTS
earlier in this study. In this case the number of `effective stars'
are calculated as the integrated overall detection probability in the
0.8--10 days period range.\footnote{For a Jupiter size planet, using
  the Kepler prior this factor is $P_\mathrm{det}=0.069$.}
\par In Fig.\ \ref{fig:kepler_frac} and Table \ref{tab:kepler_jup}
Kepler planetary fractions ($f$) are shown for short period (0.8--10
days) hot Jupiters (8-32$R_\oplus$). E.g.\ in the coolest stellar bin,
there are 9 planets (2-32$R_\oplus$) with period $T<50$ days, counting
as 320.5 occurrences around 1086 M dwarfs. This yields a fraction of
0.295 which is the same rate published in
\citetalias{Howard_2012}. This means that the correction factor for
the M dwarf bin (3600-4100K) is 1 i.e.\ all 1086 M dwarf lightcurves
are good enough to detect planets down to super-Earth sizes. As there
is no Jupiter size planet (8-32$R_\oplus$) in the 0.8--10 day period
interval, we calculate an upper limit of the planet occurrence rate as
$3/(1086\cdot 0.069)=0.04$.
\begin{table}
\begin{center}
\begin{tabular}{ccccccc} 
Temp (K) & $N_\mathrm{stars}$ & corr. & $N_\mathrm{pl}$ & $N_\mathrm{aug}$ & $f$ & $f_\mathrm{95\%}$ \\
\hline
3600--4100& 1086  & 1.00 & 0 & 0    &       & .04 \\
4100--4600& 1773  & 0.88 & 0 & 0    &       & .027  \\
4600--5100& 6029  & 0.79 & 1 & 14.5 & .003 & \\
5100--5600& 18935 & 1.00 & 6 & 55.2 & .003 & \\
5600--6100& 31407 & 1.00 & 20& 197.0& .006 & \\
6100--6600& 11808 & 0.88 & 3 & 24.9 & .002 &\\
6600--7100& 2302  & 0.76 & 1 & 3.0  & .002 & \\ 
\hline
\end{tabular}
\caption{Total number of stars ($N_\mathrm{stars}$) in Kepler Q2
  temperature bins, their corresponding correction (corr.) factors
  (see text), number of Jupiter size short period planets
  ($N_\mathrm{pl}$), their augmented contribution ($N\mathrm{aug}$)
  and the occurrence ratio ($f$) or upper limit
  ($f_\mathrm{95\%}$).\label{tab:kepler_jup}}
\end{center}
\end{table}
\par We recall that in this study 2844 early (M0-2, 3400K--3800K) and
1679 later type (M2-4, 2960K--3400K) M dwarfs were identified in the
19hr field of the WTS while there are 1086 early M dwarfs
(3600K--4100K) in the Kepler Q2 dataset. Due to the factor of 2 higher
number of early M dwarfs, the statistical upper limit set up by
present study for the WTS M0-2 bin is a stricter constraint both in
the {\em threshold} (2.9 per cent) and {\em periodmatch} approaches
(3.4 per cent) than the one that can be derived for the (early) Kepler
M dwarfs (4 per cent, Fig.\ \ref{fig:kepler_frac}). Though the WTS is
more sensitive for Jupiters around later type M dwarfs, the upper
limit for the M2-4 bin is slightly worse due to the smaller sample
size: $f_\mathrm{95\%}=$4.3--5.2 per cent (depending on the period prior), while for the
overall M0-4 sample, it is 1.7--2.0 per cent.
\par The WTS upper limits sit above the measured Hot Jupiter fractions
found for hotter stars by Kepler which have much less uncertainty from
much larger samples (bins with higher than 4600K in
Fig.\ \ref{fig:kepler_frac}). With our analysis of the WTS, we cannot
rule out a similar Hot Jupiter occurrence rate around M dwarfs
(M0-4). Our main conclusion is that there is currently no evidence to
support the argument that Hot Jupiters are less common around M dwarfs
than around solar-type stars.
\par To push our constraints to lower levels we will need to complete
the remaining three fields of the WTS, or wait for the completion of
other surveys such as Pan-Planets \citep{Koppenhoefer_2009} or PTF
\citep{Law_2012}.
\par The completed WTS survey contains a total of almost 15,000 M0-4
dwarfs. A confirmed null detection for hot Jupiters in the larger
sample would push our upper limit on the HJ fraction down by a factor
3.2, i.e. to $f_\mathrm{95\%}=$0.5--0.6 per cent for M0-4 stars. This
will then overlap with the values found for hotter stars in Fig.\
\ref{fig:kepler_frac}.
\par We note that the recent Kepler discovery and confirmation of
KOI-254 \citep{Johnson_2012}, a hot Jupiter planet around an M dwarf
is beyond the parameter space considered in the
\citetalias{Howard_2012} study thus cannot be directly incorporated
into our comparison (object is fainter). Nevertheless, if this planet
{\em were} part of the \citetalias{Howard_2012} sample, this sole
detection would mean a $1.1_{-1.1}^{+4.1}$ per cent occurrence rate
which is lower than the upper limits but still compatible with the
findings of the present and the \cite{Bonfils_2011} RV study. We note
that this is an overestimation of the real weight of this planet
detection as there are more cool hosts in KIC down to $K_p=$15.979 but
linked to unknown, possibly lower survey sensitivity as well.
\par There are some caveats in our comparison. (i) The early M dwarf
(M0-2) temperature bins are not exactly the same in the present WTS
study (3400K--3800K) and in \citetalias{Howard_2012} (3600K--4100K)
thus the WTS bins are apparently $\sim$ 250K cooler. However, we
cannot use the same method to re-estimate temperatures for Kepler
sources. Hopefully, as more follow-up data is released this will
become possible. (ii) The sample of Kepler objects used to calculate
planetary occurrences is restricted to bright objects ($K_p<15$) which
is different from the magnitude range of WTS objects studied here
($J<17$). It is possible that the Kepler early M dwarf sample is from
a different population than the sample in the present study
\citep{Mann_2012}. (iii) The sensitivity for Jupiters in the WTS was
determined for a conservative $10R_\oplus$ radius while hot Jupiters
may have larger radii. The Kepler study uses the 8--32$R_\oplus$
radius range for hot Jupiters.
\subsection{Comparison with RV studies}
\par \cite{Bonfils_2011} analyze RV data observed by the ESO/HARPS
spectrograph for planet signatures in their M dwarf (M0-6) sample of
102 nearby (<11pc, $V<14$) stars and find no Hot Jupiters. According
to their sensitivity considerations, this sample is equivalent to
96.83 effective stars. Using our 95 per cent confidence level, their
null detection of hot Jupiters (1-10 days) implies a 3.1 per cent
upper limit on the planetary occurrence rate around M dwarfs, in good
agreement with our WTS results.
\par \cite{Wright_2012} find that 1.2$\pm 0.38$ per cent of nearby
Solar-type stars host hot Jupiters. Their results are in good
agreement with previous RV works
\citep{Marcy_2005,Cumming_2008,Mayor_2011}. Occurrence rates found by
transit studies are systematically lower (\cite{Gould_2006},
\citetalias{Howard_2012}), around 0.5 per cent. 
\par As the rates around G dwarfs are about the same (RV) or lower
(transit) than the upper limits reached by the WTS around M dwarfs in
present study, we cannot rule out planet formation scenarios in this
paper. We admit that this is a very brief comparison only, avoiding
the discussion of the methods and error levels used by the studies
cited above. 
\section{Summary}
The WFCAM Transit Survey has observed 950 epochs for one of its four
target fields. These data were collected over more than three years
and will provide a valuable resource for general studies of the
photometric and astrometric properties of large numbers of objects in
the near-infrared. In this paper we determined the sensitivity of this
dataset to short period ($<$10 day) Jupiter and Neptune sized planets
around host stars of spectral type M0-M4. We identify and classify two
subsamples of M dwarfs: M0-M2 comprising 2844 stars, and M2-M4
comprising 1679 stars. We compare Dartmouth and NextGen models to
derive estimates of $T_\mathrm{eff}$ for the sample, and demonstrate
that reddening effects are small for the 19-hour field. This forms
one of the largest samples of M dwarfs targeted by any dedicated
transit survey, and currently the only one working in the
near-infrared.
\par We have described how our multi epoch WFCAM data are filtered and
processed to produce lightcurves with median $RMS\sim 4$ mmag down to
$J=14$ and $RMS\sim 17.5$ mmag at $J=17$. We found the $RMS$ for
the brightest stars suffers a significant contribution from systematic
noise at the $\geq 3$ mmag level. The origin of this is not
determined, but we note that we see similar (albeit smaller) effects
in the optical \citep[e.g.\ ][]{Irwin_2007}. We leave for future work
an investigation of the flatfield and near-infrared background as two
potential contributors to the systematics.
\par We performed Monte-Carlo simulations on the WTS lightcurves,
injecting and recovering fake transit events. We generalise 
each of the stellar and planet populations into two distinct
radius regimes, giving us 4 scenarios for transit depths: Jupiters and
Neptunes, around M0-M2 stars and M2-M4 stars. We investigate the
resultant signal-to-noise of the recovered events and compare this to
the survey detection thresholds. We also investigate our sensitivity
to these 4 scenarios as a function of orbital period and stellar
magnitude. Our analysis of the simulations enables us to place
constraints on the incidence of hot Jupiters.
\par With 95 per cent confidence and for periods $<$10 days, we showed
that fewer than 4.3--5.2 per cent of M2-M4 dwarfs host hot
Jupiters. Constraints are even stronger for earlier spectral types,
and fewer than 2.9--3.4 per cent of M0-M2 dwarfs host hot Jupiters,
while for the overall M0-4 sample, it is 1.7--2.0 per cent. An
analysis of the Kepler Q2 data shows that the WTS provides more
rigorous upper limits around cooler objects, thanks to the larger size
of our sample (2844 M dwarfs in the WTS compared to 1086 in the Kepler
sample under consideration). We compare these upper limits to the
measured hot Jupiter fraction around more massive host stars, and find
that they are consistent, i.e.\ we cannot rule out similar planet
formation scenarios around at least the earlier M dwarfs (M0-4) at the
moment.
\par We used our simulations to demonstrate that the WTS lightcurves
are sensitive to transits induced by hot Neptunes in favorable cases,
but that the ability to distinguish them from false alarms
(astrophysical or systematic) is limited by a number of
factors. Firstly the transit events have significantly lower to
signal-to-noise than events arising from hot Jupiters, and thus the
contamination by false alarms (from systematic noise) is higher at the
required detection thresholds. Secondly, the recovered periods for
the simulated hot Neptunes are dominated by spikes at multiples of 1
day, not seen so strongly for the larger planets. Without reliable
periods, any attempts at follow-up photometry and spectroscopy for
detected candidates will be hardly viable, especially for such
faint target stars.
\par The data presented in this paper represent one quarter of the
planned WFCAM Transit Survey (the other 3 fields currently lack
sufficient coverage to include in the analysis). If the WTS is
completed then we expect our statistical constraints to improve by a
factor of 3--4 (dependent on the actual numbers of M dwarfs in the
remaining three fields). If no hot Jupiters are found, our upper
limits get to the 0.5 per cent level for M0-M4 dwarfs, and may
ultimately reach a significantly lower level for hot Jupiter
occurrence than is measured around G dwarfs by other surveys.
\section{Acknowledgments}
G.K.\ and B.S.\ are supported by RoPACS, a Marie Curie Initial
Training Network funded by the European Commission's Seventh Framework
Programme. S.T.H.\ and D.P.\ are grateful to receive financial support
from RoPACS during this research. The authors are thankful to S.\
Aigrain and J.\ Irwin for their algorithms and software development
legacy from the Monitor project. This work also relies on image
processing algorithms and implementations created by former and
present members of the CASU group. The authors also thank M.\ Irwin
for his useful comments on the paper.
%
\bibliography{simpaper.bib}

\begin{thebibliography}{73}
\expandafter\ifx\csname natexlab\endcsname\relax\def\natexlab#1{#1}\fi

\bibitem[{{Abazajian} {et~al}\mbox{.}(2009){Abazajian}, {Adelman-McCarthy},
  {Ag{\"u}eros}, {Allam}, {Allende Prieto}, {An}, {Anderson}, {Anderson},
  {Annis}, {Bahcall}, \& et~al.}]{SDSS_DR7}
{Abazajian} K.~N. {et~al.}, 2009, \apjs, 182, 543

\bibitem[{{Aigrain} \& {Irwin}(2004)}]{Aigrain_Irwin_2004}
{Aigrain} S., {Irwin} M., 2004, \mnras, 350, 331

\bibitem[{{Allard}(1990)}]{Allard_1990}
{Allard} F., 1990, PhD thesis, PhD thesis.~Ruprecht Karls Univ.~Heidelberg,
  (1990)

\bibitem[{{Baraffe} \& {Chabrier}(1996)}]{Baraffe_1996}
{Baraffe} I., {Chabrier} G., 1996, \apjl, 461, L51

\bibitem[{{Baraffe} {et~al}\mbox{.}(1998){Baraffe}, {Chabrier}, {Allard}, \&
  {Hauschildt}}]{Baraffe_1998}
{Baraffe} I., {Chabrier} G., {Allard} F., {Hauschildt} P.~H., 1998, \aap, 337,
  403

\bibitem[{{Batista} {et~al}\mbox{.}(2011){Batista}, {Gould}, {Dieters}, {Dong},
  {Bond}, {Beaulieu}, {Maoz}, {Monard}, {Christie}, {McCormick}, {Albrow},
  {Horne}, {Tsapras}, {Burgdorf}, {Calchi Novati}, {Skottfelt}, {Caldwell},
  {Koz{\l}owski}, {Kubas}, {Gaudi}, {Han}, {Bennett}, {An}, {The MOA
  Collaboration}, {Abe}, {Botzler}, {Douchin}, {Freeman}, {Fukui}, {Furusawa},
  {Hearnshaw}, {Hosaka}, {Itow}, {Kamiya}, {Kilmartin}, {Korpela}, {Lin},
  {Ling}, {Makita}, {Masuda}, {Matsubara}, {Miyake}, {Muraki}, {Nagaya},
  {Nishimoto}, {Ohnishi}, {Okumura}, {Perrott}, {Rattenbury}, {Saito},
  {Sullivan}, {Sumi}, {Sweatman}, {Tristram}, {von Seggern}, {Yock}, {The
  PLANET Collaboration}, {Brillant}, {Calitz}, {Cassan}, {Cole}, {Cook},
  {Coutures}, {Dominis Prester}, {Donatowicz}, {Greenhill}, {Hoffman},
  {Jablonski}, {Kane}, {Kains}, {Marquette}, {Martin}, {Martioli}, {Meintjes},
  {Menzies}, {Pedretti}, {Pollard}, {Sahu}, {Vinter}, {Wambsganss}, {Watson},
  {Williams}, {Zub}, {The FUN Collaboration}, {Allen}, {Bolt}, {Bos}, {DePoy},
  {Drummond}, {Eastman}, {Gal-Yam}, {Gorbikov}, {Higgins}, {Janczak}, {Kaspi},
  {Lee}, {Mallia}, {Maury}, {Monard}, {Moorhouse}, {Morgan}, {Natusch}, {Ofek},
  {Park}, {Pogge}, {Polishook}, {Santallo}, {Shporer}, {Spector}, {Thornley},
  {Yee}, {The MiNDSTEp Consortium}, {Bozza}, {Browne}, {Dominik}, {Dreizler},
  {Finet}, {Glitrup}, {Grundahl}, {Harps{\o}e}, {Hessman}, {Hinse},
  {Hundertmark}, {J{\o}rgensen}, {Liebig}, {Maier}, {Mancini}, {Mathiasen},
  {Rahvar}, {Ricci}, {Scarpetta}, {Southworth}, {Surdej}, {Zimmer}, {The
  RoboNet Collaboration}, {Allan}, {Bramich}, {Snodgrass}, {Steele}, \&
  {Street}}]{Batista_2011}
{Batista} V. {et~al.}, 2011, \aap, 529, A102+

\bibitem[{{Berta} {et~al}\mbox{.}(2012){Berta}, {Irwin}, {Charbonneau},
  {Burke}, \& {Falco}}]{Berta_2012}
{Berta} Z.~K., {Irwin} J., {Charbonneau} D., {Burke} C.~J., {Falco} E.~E.,
  2012, \aj, 144, 145

\bibitem[{Birkby {et~al}\mbox{.}(2013)Birkby {et~al.}}]{Birkby_2013}
Birkby J., {et~al.}, 2013, in prep.

\bibitem[{{Bonfils} {et~al}\mbox{.}(2011){Bonfils}, {Delfosse}, {Udry},
  {Forveille}, {Mayor}, {Perrier}, {Bouchy}, {Gillon}, {Lovis}, {Pepe},
  {Queloz}, {Santos}, {S{\'e}gransan}, \& {Bertaux}}]{Bonfils_2011}
{Bonfils} X. {et~al.}, 2011, ArXiv e-prints

\bibitem[{{Borucki} {et~al}\mbox{.}(2011){Borucki}, {Koch}, {Basri}, {Batalha},
  {Brown}, {Bryson}, {Caldwell}, {Christensen-Dalsgaard}, {Cochran}, {DeVore},
  {Dunham}, {Gautier}, {Geary}, {Gilliland}, {Gould}, {Howell}, {Jenkins},
  {Latham}, {Lissauer}, {Marcy}, {Rowe}, {Sasselov}, {Boss}, {Charbonneau},
  {Ciardi}, {Doyle}, {Dupree}, {Ford}, {Fortney}, {Holman}, {Seager},
  {Steffen}, {Tarter}, {Welsh}, {Allen}, {Buchhave}, {Christiansen}, {Clarke},
  {D{\'e}sert}, {Endl}, {Fabrycky}, {Fressin}, {Haas}, {Horch}, {Howard},
  {Isaacson}, {Kjeldsen}, {Kolodziejczak}, {Kulesa}, {Li}, {Machalek},
  {McCarthy}, {MacQueen}, {Meibom}, {Miquel}, {Prsa}, {Quinn}, {Quintana},
  {Ragozzine}, {Sherry}, {Shporer}, {Tenenbaum}, {Torres}, {Twicken}, {Van
  Cleve}, \& {Walkowicz}}]{Borucki_2011}
{Borucki} W.~J. {et~al.}, 2011, ArXiv e-prints

\bibitem[{{Borucki} {et~al}\mbox{.}(1997){Borucki}, {Koch}, {Dunham}, \&
  {Jenkins}}]{Kepler}
{Borucki} W.~J., {Koch} D.~G., {Dunham} E.~W., {Jenkins} J.~M., 1997, in
  Astronomical Society of the Pacific Conference Series, Vol. 119, Planets
  Beyond the Solar System and the Next Generation of Space Missions,
  {D.~Soderblom}, ed., p. 153

\bibitem[{{Boss}(2006)}]{Boss_2006}
{Boss} A.~P., 2006, \apj, 643, 501

\bibitem[{{Brown} {et~al}\mbox{.}(2011){Brown}, {Latham}, {Everett}, \&
  {Esquerdo}}]{Brown_2011}
{Brown} T.~M., {Latham} D.~W., {Everett} M.~E., {Esquerdo} G.~A., 2011, \aj,
  142, 112

\bibitem[{{Burke} {et~al}\mbox{.}(2006){Burke}, {Gaudi}, {DePoy}, \&
  {Pogge}}]{Burke_2006}
{Burke} C.~J., {Gaudi} B.~S., {DePoy} D.~L., {Pogge} R.~W., 2006, \aj, 132, 210

\bibitem[{{Calabretta} \& {Greisen}(2002)}]{Calabretta_2002}
{Calabretta} M.~R., {Greisen} E.~W., 2002, \aap, 395, 1077

\bibitem[{{Cappetta} {et~al}\mbox{.}(2012){Cappetta}, {Saglia}, {Birkby},
  {Koppenhoefer}, {Pinfield}, {Hodgkin}, {Cruz}, {Kov{\'a}cs}, {Sip{\"o}cz},
  {Barrado}, {Nefs}, {Pavlenko}, {Fossati}, {del Burgo}, {Mart{\'{\i}}n},
  {Snellen}, {Barnes}, {Bayo}, {Campbell}, {Catalan}, {G{\'a}lvez-Ortiz},
  {Goulding}, {Haswell}, {Ivanyuk}, {Jones}, {Kuznetsov}, {Lodieu}, {Marocco},
  {Mislis}, {Murgas}, {Napiwotzki}, {Palle}, {Pollacco}, {Sarro Baro},
  {Solano}, {Steele}, {Stoev}, {Tata}, \& {Zendejas}}]{Cappetta_2012}
{Cappetta} M. {et~al.}, 2012, ArXiv e-prints

\bibitem[{{Cardelli}, {Clayton} \& {Mathis}(1989){Cardelli}, {Clayton}, \&
  {Mathis}}]{Cardelli_1989}
{Cardelli} J.~A., {Clayton} G.~C., {Mathis} J.~S., 1989, \apj, 345, 245

\bibitem[{{Castelli} \& {Kurucz}(2004)}]{CK_2004}
{Castelli} F., {Kurucz} R.~L., 2004, ArXiv Astrophysics e-prints

\bibitem[{{Chabrier}(2003)}]{Chabrier_2003}
{Chabrier} G., 2003, \pasp, 115, 763

\bibitem[{{Chabrier}, {Gallardo} \& {Baraffe}(2007){Chabrier}, {Gallardo}, \&
  {Baraffe}}]{Chabrier_2007}
{Chabrier} G., {Gallardo} J., {Baraffe} I., 2007, \aap, 472, L17

\bibitem[{{Charbonneau} {et~al}\mbox{.}(2009){Charbonneau}, {Berta}, {Irwin},
  {Burke}, {Nutzman}, {Buchhave}, {Lovis}, {Bonfils}, {Latham}, {Udry},
  {Murray-Clay}, {Holman}, {Falco}, {Winn}, {Queloz}, {Pepe}, {Mayor},
  {Delfosse}, \& {Forveille}}]{Charbonneau_2009}
{Charbonneau} D. {et~al.}, 2009, \nat, 462, 891

\bibitem[{{Chazelas} {et~al}\mbox{.}(2012){Chazelas}, {Pollacco}, {Queloz},
  {Rauer}, {Wheatley}, {West}, {Da Silva Bento}, {Burleigh}, {McCormac},
  {Eigm{\"u}ller}, {Erikson}, {Genolet}, {Goad}, {Jord{\'a}n}, {Neveu}, \&
  {Walker}}]{Chazelas_2012}
{Chazelas} B. {et~al.}, 2012, in Society of Photo-Optical Instrumentation
  Engineers (SPIE) Conference Series, Vol. 8444, Society of Photo-Optical
  Instrumentation Engineers (SPIE) Conference Series

\bibitem[{{Ciardi} {et~al}\mbox{.}(2011){Ciardi}, {von Braun}, {Bryden}, {van
  Eyken}, {Howell}, {Kane}, {Plavchan}, {Ram{\'{\i}}rez}, \&
  {Stauffer}}]{Ciardi_2011}
{Ciardi} D.~R. {et~al.}, 2011, \aj, 141, 108

\bibitem[{{Claret}(2000)}]{Claret_2000}
{Claret} A., 2000, \aap, 363, 1081

\bibitem[{{Covey} {et~al}\mbox{.}(2007){Covey}, {Ivezi{\'c}}, {Schlegel},
  {Finkbeiner}, {Padmanabhan}, {Lupton}, {Ag{\"u}eros}, {Bochanski}, {Hawley},
  {West}, {Seth}, {Kimball}, {Gogarten}, {Claire}, {Haggard}, {Kaib},
  {Schneider}, \& {Sesar}}]{Covey_2007}
{Covey} K.~R. {et~al.}, 2007, \aj, 134, 2398

\bibitem[{{Cumming} {et~al}\mbox{.}(2008){Cumming}, {Butler}, {Marcy}, {Vogt},
  {Wright}, \& {Fischer}}]{Cumming_2008}
{Cumming} A., {Butler} R.~P., {Marcy} G.~W., {Vogt} S.~S., {Wright} J.~T.,
  {Fischer} D.~A., 2008, \pasp, 120, 531

\bibitem[{{Dotter} {et~al}\mbox{.}(2008){Dotter}, {Chaboyer}, {Jevremovi{\'c}},
  {Kostov}, {Baron}, \& {Ferguson}}]{Dotter_2008}
{Dotter} A., {Chaboyer} B., {Jevremovi{\'c}} D., {Kostov} V., {Baron} E.,
  {Ferguson} J.~W., 2008, \apjs, 178, 89

\bibitem[{{Drimmel}, {Cabrera-Lavers} \& {L{\'o}pez-Corredoira}(2003){Drimmel},
  {Cabrera-Lavers}, \& {L{\'o}pez-Corredoira}}]{Drimmel_2003}
{Drimmel} R., {Cabrera-Lavers} A., {L{\'o}pez-Corredoira} M., 2003, \aap, 409,
  205

\bibitem[{{Dzigan} \& {Zucker}(2011)}]{Dzigan_2011}
{Dzigan} Y., {Zucker} S., 2011, \mnras, 415, 2513

\bibitem[{{Fasano} \& {Franceschini}(1987)}]{KS_2D}
{Fasano} G., {Franceschini} A., 1987, \mnras, 225, 155

\bibitem[{{Gillon} {et~al}\mbox{.}(2007){Gillon}, {Pont}, {Demory}, {Mallmann},
  {Mayor}, {Mazeh}, {Queloz}, {Shporer}, {Udry}, \& {Vuissoz}}]{Gillon_2007}
{Gillon} M. {et~al.}, 2007, \aap, 472, L13

\bibitem[{{Gould} {et~al}\mbox{.}(2010){Gould}, {Dong}, {Gaudi}, {Udalski},
  {Bond}, {Greenhill}, {Street}, {Dominik}, {Sumi}, {Szyma{\'n}ski}, {Han},
  {Allen}, {Bolt}, {Bos}, {Christie}, {DePoy}, {Drummond}, {Eastman},
  {Gal-Yam}, {Higgins}, {Janczak}, {Kaspi}, {Koz{\l}owski}, {Lee}, {Mallia},
  {Maury}, {Maoz}, {McCormick}, {Monard}, {Moorhouse}, {Morgan}, {Natusch},
  {Ofek}, {Park}, {Pogge}, {Polishook}, {Santallo}, {Shporer}, {Spector},
  {Thornley}, {Yee}, {{$\mu$}FUN Collaboration}, {Kubiak}, {Pietrzy{\'n}ski},
  {Soszy{\'n}ski}, {Szewczyk}, {Wyrzykowski}, {Ulaczyk}, {Poleski}, {OGLE
  Collaboration}, {Abe}, {Bennett}, {Botzler}, {Douchin}, {Freeman}, {Fukui},
  {Furusawa}, {Hearnshaw}, {Hosaka}, {Itow}, {Kamiya}, {Kilmartin}, {Korpela},
  {Lin}, {Ling}, {Makita}, {Masuda}, {Matsubara}, {Miyake}, {Muraki}, {Nagaya},
  {Nishimoto}, {Ohnishi}, {Okumura}, {Perrott}, {Philpott}, {Rattenbury},
  {Saito}, {Sako}, {Sullivan}, {Sweatman}, {Tristram}, {von Seggern}, {Yock},
  {MOA Collaboration}, {Albrow}, {Batista}, {Beaulieu}, {Brillant}, {Caldwell},
  {Calitz}, {Cassan}, {Cole}, {Cook}, {Coutures}, {Dieters}, {Dominis Prester},
  {Donatowicz}, {Fouqu{\'e}}, {Hill}, {Hoffman}, {Jablonski}, {Kane}, {Kains},
  {Kubas}, {Marquette}, {Martin}, {Martioli}, {Meintjes}, {Menzies},
  {Pedretti}, {Pollard}, {Sahu}, {Vinter}, {Wambsganss}, {Watson}, {Williams},
  {Zub}, {PLANET Collaboration}, {Allan}, {Bode}, {Bramich}, {Burgdorf},
  {Clay}, {Fraser}, {Hawkins}, {Horne}, {Kerins}, {Lister}, {Mottram},
  {Saunders}, {Snodgrass}, {Steele}, {Tsapras}, {RoboNet Collaboration},
  {J{\o}rgensen}, {Anguita}, {Bozza}, {Calchi Novati}, {Harps{\o}e}, {Hinse},
  {Hundertmark}, {Kj{\ae}rgaard}, {Liebig}, {Mancini}, {Masi}, {Mathiasen},
  {Rahvar}, {Ricci}, {Scarpetta}, {Southworth}, {Surdej}, {Th{\"o}ne}, \&
  {MiNDSTEp Consortium}}]{Gould_2010}
{Gould} A. {et~al.}, 2010, \apj, 720, 1073

\bibitem[{{Gould} {et~al}\mbox{.}(2006){Gould}, {Dorsher}, {Gaudi}, \&
  {Udalski}}]{Gould_2006}
{Gould} A., {Dorsher} S., {Gaudi} B.~S., {Udalski} A., 2006, \actaa, 56, 1

\bibitem[{{Goulding} {et~al}\mbox{.}(2012){Goulding}, {Barnes}, {Pinfield},
  {Kov{\'a}cs}, {Birkby}, {Hodgkin}, {Catal{\'a}n}, {Sip{\H o}cz}, {Jones},
  {del Burgo}, {Jeffers}, {Nefs}, {G{\'a}lvez-Ortiz}, \&
  {Martin}}]{Goulding_2011}
{Goulding} N.~T. {et~al.}, 2012, ArXiv e-prints

\bibitem[{{Greisen} \& {Calabretta}(2002)}]{Greisen_2002}
{Greisen} E.~W., {Calabretta} M.~R., 2002, \aap, 395, 1061

\bibitem[{{Hartman} {et~al}\mbox{.}(2009){Hartman}, {Gaudi}, {Holman},
  {McLeod}, {Stanek}, {Barranco}, {Pinsonneault}, {Meibom}, \&
  {Kalirai}}]{Hartman_2009}
{Hartman} J.~D. {et~al.}, 2009, \apj, 695, 336

\bibitem[{{Hewett} {et~al}\mbox{.}(2006){Hewett}, {Warren}, {Leggett}, \&
  {Hodgkin}}]{Hewett_2006}
{Hewett} P.~C., {Warren} S.~J., {Leggett} S.~K., {Hodgkin} S.~T., 2006, \mnras,
  367, 454

\bibitem[{{Hodgkin} {et~al}\mbox{.}(2009){Hodgkin}, {Irwin}, {Hewett}, \&
  {Warren}}]{Hodgkin_2009}
{Hodgkin} S.~T., {Irwin} M.~J., {Hewett} P.~C., {Warren} S.~J., 2009, \mnras,
  394, 675

\bibitem[{{Howard} {et~al}\mbox{.}(2012){Howard}, {Marcy}, {Bryson}, {Jenkins},
  {Rowe}, {Batalha}, {Borucki}, {Koch}, {Dunham}, {Gautier}, {Van Cleve},
  {Cochran}, {Latham}, {Lissauer}, {Torres}, {Brown}, {Gilliland}, {Buchhave},
  {Caldwell}, {Christensen-Dalsgaard}, {Ciardi}, {Fressin}, {Haas}, {Howell},
  {Kjeldsen}, {Seager}, {Rogers}, {Sasselov}, {Steffen}, {Basri},
  {Charbonneau}, {Christiansen}, {Clarke}, {Dupree}, {Fabrycky}, {Fischer},
  {Ford}, {Fortney}, {Tarter}, {Girouard}, {Holman}, {Johnson}, {Klaus},
  {Machalek}, {Moorhead}, {Morehead}, {Ragozzine}, {Tenenbaum}, {Twicken},
  {Quinn}, {Isaacson}, {Shporer}, {Lucas}, {Walkowicz}, {Welsh}, {Boss},
  {Devore}, {Gould}, {Smith}, {Morris}, {Prsa}, {Morton}, {Still}, {Thompson},
  {Mullally}, {Endl}, \& {MacQueen}}]{Howard_2012}
{Howard} A.~W. {et~al.}, 2012, \apjs, 201, 15

\bibitem[{{Ida} \& {Lin}(2005)}]{Ida_Lin_2005}
{Ida} S., {Lin} D.~N.~C., 2005, \apj, 626, 1045

\bibitem[{{Ida} \& {Lin}(2008)}]{Ida_Lin_2008}
{Ida} S., {Lin} D.~N.~C., 2008, \apj, 685, 584

\bibitem[{{Irwin} {et~al}\mbox{.}(2007){Irwin}, {Irwin}, {Aigrain}, {Hodgkin},
  {Hebb}, \& {Moraux}}]{Irwin_2007}
{Irwin} J., {Irwin} M., {Aigrain} S., {Hodgkin} S., {Hebb} L., {Moraux} E.,
  2007, \mnras, 375, 1449

\bibitem[{{Irwin} \& {Lewis}(2001)}]{Irwin_Lewis_2001}
{Irwin} M., {Lewis} J., 2001, \nar, 45, 105

\bibitem[{{Irwin}(1985)}]{Irwin_1985}
{Irwin} M.~J., 1985, \mnras, 214, 575

\bibitem[{{Irwin} {et~al}\mbox{.}(2004){Irwin}, {Lewis}, {Hodgkin}, {Bunclark},
  {Evans}, {McMahon}, {Emerson}, {Stewart}, \& {Beard}}]{Irwin_2004}
{Irwin} M.~J. {et~al.}, 2004, in Society of Photo-Optical Instrumentation
  Engineers (SPIE) Conference Series, Vol. 5493, Society of Photo-Optical
  Instrumentation Engineers (SPIE) Conference Series, {Quinn} P.~J., {Bridger}
  A., eds., pp. 411--422

\bibitem[{{Johnson} {et~al}\mbox{.}(2010){Johnson}, {Aller}, {Howard}, \&
  {Crepp}}]{Johnson_2010}
{Johnson} J.~A., {Aller} K.~M., {Howard} A.~W., {Crepp} J.~R., 2010, \pasp,
  122, 905

\bibitem[{{Johnson} {et~al}\mbox{.}(2007){Johnson}, {Butler}, {Marcy},
  {Fischer}, {Vogt}, {Wright}, \& {Peek}}]{Johnson_2007}
{Johnson} J.~A., {Butler} R.~P., {Marcy} G.~W., {Fischer} D.~A., {Vogt} S.~S.,
  {Wright} J.~T., {Peek} K.~M.~G., 2007, \apj, 670, 833

\bibitem[{{Johnson} {et~al}\mbox{.}(2012){Johnson}, {Gazak}, {Apps},
  {Muirhead}, {Crepp}, {Crossfield}, {Boyajian}, {von Braun}, {Rojas-Ayala},
  {Howard}, {Covey}, {Schlawin}, {Hamren}, {Morton}, {Marcy}, \&
  {Lloyd}}]{Johnson_2012}
{Johnson} J.~A. {et~al.}, 2012, \aj, 143, 111

\bibitem[{{Kennedy} \& {Kenyon}(2008)}]{Kennedy_Kenyon_2008}
{Kennedy} G.~M., {Kenyon} S.~J., 2008, \apj, 673, 502

\bibitem[{{Koppenhoefer} {et~al}\mbox{.}(2009){Koppenhoefer}, {Afonso},
  {Saglia}, \& {Henning}}]{Koppenhoefer_2009}
{Koppenhoefer} J., {Afonso} C., {Saglia} R.~P., {Henning} T., 2009, \aap, 494,
  707

\bibitem[{{Kov{\'a}cs}, {Zucker} \& {Mazeh}(2002){Kov{\'a}cs}, {Zucker}, \&
  {Mazeh}}]{Kovacs_2002}
{Kov{\'a}cs} G., {Zucker} S., {Mazeh} T., 2002, \aap, 391, 369

\bibitem[{{Kraus} {et~al}\mbox{.}(2011){Kraus}, {Tucker}, {Thompson}, {Craine},
  \& {Hillenbrand}}]{Kraus_2011}
{Kraus} A.~L., {Tucker} R.~A., {Thompson} M.~I., {Craine} E.~R., {Hillenbrand}
  L.~A., 2011, \apj, 728, 48

\bibitem[{{Laughlin}, {Bodenheimer} \& {Adams}(2004){Laughlin}, {Bodenheimer},
  \& {Adams}}]{Laughlin_2004}
{Laughlin} G., {Bodenheimer} P., {Adams} F.~C., 2004, \apjl, 612, L73

\bibitem[{{Law} {et~al}\mbox{.}(2012){Law}, {Kraus}, {Street}, {Fulton},
  {Hillenbrand}, {Shporer}, {Lister}, {Baranec}, {Bloom}, {Bui}, {Burse},
  {Cenko}, {Das}, {Davis}, {Dekany}, {Filippenko}, {Kasliwal}, {Kulkarni},
  {Nugent}, {Ofek}, {Poznanski}, {Quimby}, {Ramaprakash}, {Riddle},
  {Silverman}, {Sivanandam}, \& {Tendulkar}}]{Law_2012}
{Law} N.~M. {et~al.}, 2012, \apj, 757, 133

\bibitem[{{Lawrence} {et~al}\mbox{.}(2007){Lawrence}, {Warren}, {Almaini},
  {Edge}, {Hambly}, {Jameson}, {Lucas}, {Casali}, {Adamson}, {Dye}, {Emerson},
  {Foucaud}, {Hewett}, {Hirst}, {Hodgkin}, {Irwin}, {Lodieu}, {McMahon},
  {Simpson}, {Smail}, {Mortlock}, \& {Folger}}]{Lawrence_2007}
{Lawrence} A. {et~al.}, 2007, \mnras, 379, 1599

\bibitem[{{Leggett}(1992)}]{Leggett_1992}
{Leggett} S.~K., 1992, \apjs, 82, 351

\bibitem[{{Mandel} \& {Agol}(2002)}]{Mandel_Agol_2002}
{Mandel} K., {Agol} E., 2002, \apjl, 580, L171

\bibitem[{{Mann} {et~al}\mbox{.}(2012){Mann}, {Gaidos}, {L{\'e}pine}, \&
  {Hilton}}]{Mann_2012}
{Mann} A.~W., {Gaidos} E., {L{\'e}pine} S., {Hilton} E., 2012, ArXiv e-prints

\bibitem[{{Marcy} {et~al}\mbox{.}(2005){Marcy}, {Butler}, {Fischer}, {Vogt},
  {Wright}, {Tinney}, \& {Jones}}]{Marcy_2005}
{Marcy} G., {Butler} R.~P., {Fischer} D., {Vogt} S., {Wright} J.~T., {Tinney}
  C.~G., {Jones} H.~R.~A., 2005, Progress of Theoretical Physics Supplement,
  158, 24

\bibitem[{{Mayor} {et~al}\mbox{.}(2011){Mayor}, {Marmier}, {Lovis}, {Udry},
  {S{\'e}gransan}, {Pepe}, {Benz}, {Bertaux}, {Bouchy}, {Dumusque}, {Lo Curto},
  {Mordasini}, {Queloz}, \& {Santos}}]{Mayor_2011}
{Mayor} M. {et~al.}, 2011, ArXiv e-prints

\bibitem[{{Miller} {et~al}\mbox{.}(2008){Miller}, {Irwin}, {Aigrain},
  {Hodgkin}, \& {Hebb}}]{Miller_2008}
{Miller} A.~A., {Irwin} J., {Aigrain} S., {Hodgkin} S., {Hebb} L., 2008,
  \mnras, 387, 349

\bibitem[{{Mould}(1976)}]{Mould_1976}
{Mould} J.~R., 1976, \aap, 48, 443

\bibitem[{{O'Donnell}(1994)}]{ODonnell_1994}
{O'Donnell} J.~E., 1994, \apj, 422, 158

\bibitem[{{P{\'a}l}(2008)}]{Pal_2008}
{P{\'a}l} A., 2008, \mnras, 390, 281

\bibitem[{{Pont}, {Zucker} \& {Queloz}(2006){Pont}, {Zucker}, \&
  {Queloz}}]{Pont_2006}
{Pont} F., {Zucker} S., {Queloz} D., 2006, \mnras, 373, 231

\bibitem[{{Rodler} {et~al}\mbox{.}(2012){Rodler}, {Deshpande}, {Zapatero
  Osorio}, {Mart{\'{\i}}n}, {Montgomery}, {Del Burgo}, \&
  {Creevey}}]{Rodler_2012}
{Rodler} F., {Deshpande} R., {Zapatero Osorio} M.~R., {Mart{\'{\i}}n} E.~L.,
  {Montgomery} M.~M., {Del Burgo} C., {Creevey} O.~L., 2012, \aap, 538, A141

\bibitem[{{Schlegel}, {Finkbeiner} \& {Davis}(1998){Schlegel}, {Finkbeiner}, \&
  {Davis}}]{Schlegel_1998}
{Schlegel} D.~J., {Finkbeiner} D.~P., {Davis} M., 1998, \apj, 500, 525

\bibitem[{Sip\H{o}cz {et~al}\mbox{.}(2013)Sip\H{o}cz {et~al.}}]{Sipocz_2013}
Sip\H{o}cz B., {et~al.}, 2013, in prep.

\bibitem[{{Skrutskie} {et~al}\mbox{.}(2006){Skrutskie}, {Cutri}, {Stiening},
  {Weinberg}, {Schneider}, {Carpenter}, {Beichman}, {Capps}, {Chester},
  {Elias}, {Huchra}, {Liebert}, {Lonsdale}, {Monet}, {Price}, {Seitzer},
  {Jarrett}, {Kirkpatrick}, {Gizis}, {Howard}, {Evans}, {Fowler}, {Fullmer},
  {Hurt}, {Light}, {Kopan}, {Marsh}, {McCallon}, {Tam}, {Van Dyk}, \&
  {Wheelock}}]{2MASS}
{Skrutskie} M.~F. {et~al.}, 2006, \aj, 131, 1163

\bibitem[{{Thommes}, {Matsumura} \& {Rasio}(2008){Thommes}, {Matsumura}, \&
  {Rasio}}]{Thommes_2008}
{Thommes} E.~W., {Matsumura} S., {Rasio} F.~A., 2008, Science, 321, 814

\bibitem[{{West} {et~al}\mbox{.}(2011){West}, {Morgan}, {Bochanski},
  {Andersen}, {Bell}, {Kowalski}, {Davenport}, {Hawley}, {Schmidt}, {Bernat},
  {Hilton}, {Muirhead}, {Covey}, {Rojas-Ayala}, {Schlawin}, {Gooding},
  {Schluns}, {Dhital}, {Pineda}, \& {Jones}}]{West_2011}
{West} A.~A. {et~al.}, 2011, \aj, 141, 97

\bibitem[{{Wright} {et~al}\mbox{.}(2012){Wright}, {Marcy}, {Howard}, {Johnson},
  {Morton}, \& {Fischer}}]{Wright_2012}
{Wright} J.~T., {Marcy} G.~W., {Howard} A.~W., {Johnson} J.~A., {Morton} T.~D.,
  {Fischer} D.~A., 2012, \apj, 753, 160

\bibitem[{Zendejas {et~al}\mbox{.}(2013)Zendejas {et~al.}}]{Zendejas_2013}
Zendejas J., {et~al.}, 2013, in prep.

\end{thebibliography}
\end{document}